\documentclass[preprint,12pt,authoryear, review]{elsarticle}

\usepackage{svg}
\usepackage{gensymb}
\usepackage{amssymb}
\usepackage{amsmath}
\usepackage{array}
\usepackage{booktabs}
\usepackage{multirow}
\usepackage{caption}
\usepackage{subcaption}
\usepackage{float}
\usepackage{cleveref}
\usepackage{tabularray}
\usepackage{graphicx}
\usepackage{vcell}
\usepackage[margin=1.25in]{geometry}
\usepackage{setspace}
\usepackage[final]{changes}
\definechangesauthor[color=purple, name={Yongjin Choi}]{YC}
\usepackage{url}

\usepackage{amsthm}

\usepackage{lineno}
\usepackage{setspace}

\begin{document}
\begin{frontmatter}



\title{Graph Neural Network-based surrogate model for granular flows}


\author[label1]{Yongjin Choi\corref{cor1}}
\cortext[cor1]{Corresponding author}
\ead{yj.choi@utexas.edu}
\author[label1]{Krishna Kumar}
\affiliation[label1]{organization={Department of Civil Architectural and Environmental Engineering Austin, The University of Texas at Austin},
            city={Austin},
            postcode={78712}, 
            state={TX},
            country={USA}}

\begin{abstract}
Accurate simulation of granular flow dynamics is crucial for assessing geotechnical risks, including landslides and debris flows. Traditional numerical methods are limited by their computational cost in simulating large-scale systems. Statistical or machine learning-based models offer alternatives. Still, they are largely empirical, based on limited parameters. Due to their permutation-dependent learning, traditional machine learning-based models require huge training data to generalize. To resolve these problems, we use a graph neural network (GNN), a state-of-the-art machine learning architecture that learns local interactions. Graphs represent the state of dynamically changing granular flows and their interactions. \added{We implement a multi-Graphics Processing Units (GPU) GNN simulator (GNS) capable of handling different material types. We demonstrate the capability of GNS by modeling granular flow interactions with barriers.} GNS takes the granular flow’s current state and predicts the next state using Euler explicit integration by learning the local interaction laws. We train GNS on different granular trajectories. We then assess its performance by predicting granular column collapse and interaction with barriers. GNS accurately predicts flow dynamics for column collapses with different aspect ratios and interaction with barriers with configurations unseen during training. GNS is up to a few thousand times faster than high-fidelity numerical simulators.
\end{abstract}


\end{frontmatter}


\section{Introduction}
\label{sec:intro}

Landslides cause extensive material displacement and significant infrastructure damage. Accurate modeling of granular flow runout is crucial to understanding the impact of landslides. Numerical methods, such as particle-based and continuum approaches, are often employed to assess landslide runouts. Particle-based approaches, like the Discrete Element Method (DEM) \citep{staron2005, kermani2015, kumar2017_lbm-dem}, can model grain-grain interactions but are limited to representative elemental volumes. Traditional continuum approaches, such as the Finite Element Method, can predict the initiation of such failures but suffer from mesh distortions when capturing runout dynamics. Hybrid Eulerian-Lagrangian approaches like the Material Point Method (MPM) \citep{mast2014, kumar2017_mpm-dem} can simulate large-deformation flows without undergoing mesh distortions. However, the hybrid nature of MPM requires tracking both the grid and the material points, which is computationally expensive. Multiple full-scale simulations are necessary for a comprehensive evaluation of runout hazard scenarios. Similarly, a back analysis to estimate material parameters requires a broad parametric sweep involving hundreds to thousands of simulations. However, current state-of-the-art numerical methods are restricted to, at most, a few full-scale simulations, limiting our ability in scenario testing or back analysis.

An alternative to numerical simulations is to develop statistical or machine learning models to evaluate landslide risks. These surrogate models build correlations between landslide risks and their influencing factors. \added{However, they often oversimplify the granular flow process by relying on simple empirical correlations and statistical patterns mapped on a low-dimensional domain without considering the complex dynamics of granular flows}. Several studies adopt probabilistic approaches, such as Monte Carlo simulation and Bayesian analysis, to evaluate the landslide runout distance based on factors including topology and geology \citep{gao2021, zeng2021, sun2021, Zhao2022_bayesian_runout}. Machine learning models can predict the travel distance and potential path of granular flows based on the geometry and ground properties \citep{Durante2021, Ju2022, yang2021}. Although researchers have been able to correlate the runout of granular flow based on statistical or data-driven techniques, these techniques do not explicitly consider granular flow dynamics—the actual physics governing the flow behavior. Thus, due to a lack of physics, these models do not generalize outside their training range, such as, modeling other boundary conditions or geometry.

\added{In order to develop a surrogate model for granular flows that generalize beyond the training datasets, it is essential to understand the interaction laws governing the behavior of granular masses. Traditional machine learning models like Convolutional Neural Networks (CNNs) and Multi-Layer Perceptrons (MLPs) face difficulties learning these interactions as granular systems continuously evolve and rearrange. CNNs, while adept at learning spatially invariant features through pooling operations, are limited by their reliance on fixed, mesh-based structures. This limitation is significant in modeling granular flows where particle interactions and arrangements are neither fixed nor regular. MLPs can theoretically model dynamic systems by mapping their physical states to dynamics. However, their lack of permutation invariance, i.e., their output depends on the order of inputs, means an exponential increase in the required datasets $(O(n!))$ to map the entire parameter space of particle arrangements~\citep{Battaglia2018, haeri2022}.} 

\added{To address these limitations, we use graph neural networks (GNNs), a state-of-the-art machine learning architecture that enables permutation invariant learning~\citep{Battaglia2016, Battaglia2018, Sanchez2020}. GNNs represent the physical state of granular systems as graphs, where nodes symbolize individual particles or groups of particles, and edges depict their interactions or relational dependencies. This graph-based approach allows GNNs to learn directly from particle interactions, regardless of their spatial arrangement or sequence in the input data. It effectively captures the complex and dynamic nature of granular flows in a permutation-invariant way. GNNs adapt to structural changes in the system over time, effectively learning the underlying interaction laws.} 
 
\added{We implement a multi-Graphics Processing Units (GPU) GNN simulator (GNS) capable of handling different material types.} We demonstrate the capability of GNS by replicating the collapse of a granular column and its interaction with barriers. Granular column collapse is a simple physical experiment that captures the overall dynamics of large-scale runouts. GNS, trained on granular flow trajectories, successfully predicts the runout dynamics of column collapse and the interaction with barriers outside its training range and generalizes to upscaled domain sizes.

\section{Methods}\label{sec:method}
This section describes the individual components of GNS: graphs, graph neural networks (GNNs), and message passing.

\subsection{Graph neural networks and message passing}
\subsubsection{Graphs}
Graphs can represent interactions in physical systems \citep{Battaglia2016, Sanchez2020}. We represent the granular media as a graph $G=(\boldsymbol{V}, \boldsymbol{E})$ consisting of a set of vertices ($\boldsymbol{v}_{i} \in \boldsymbol{V}$) representing the soil grains or aggregation of grains and edges ($\boldsymbol{e}_{i,j} \in \boldsymbol{E}$) connecting a pair of vertices ($\boldsymbol{v}_i$ and $\boldsymbol{v}_j$) representing the interaction between the grains. Consider an example involving interaction between grains in a box (see \cref{fig:balls-in-box}). We encode the state of the physical system, such as the kinematics of grains and their interaction (\cref{fig:balls-in-box}a and \cref{fig:balls-in-box}d), as a graph (\cref{fig:balls-in-box}b and \cref{fig:balls-in-box}c). The vertices describe the position and velocity of the grains, and the edges describe the directional interaction between them, shown as arrows in \cref{fig:balls-in-box}b and \cref{fig:balls-in-box}c. The state of the grain $i$ is represented as a vertex feature vector $\boldsymbol{v}_i$. The vertex feature vector includes velocities, mass, and distance to the boundary. The edge feature vector $\boldsymbol{e}_{i,j}$ includes information about the interaction between grains $i$ and $j$ such as the relative distance between the grains. Thus, we can store and process the state of granular bodies and their interactions as graphs.

Graphs offer a permutation-invariant form of encoding data, where the interaction between vertices is independent of the order of vertices or their position in Euclidean space. As graphs represent the interactions between grains as edge connections, graphs are permutation invariants. For example, by storing the relative positional information in $\boldsymbol{e}_{i,j}$, rather than the absolute position, machine learning models operating on these networks learn the interaction behavior of different relative distances between grains. Therefore, graphs can efficiently represent the physical state of granular flow involving multi-grain interactions. 

\subsubsection{Graph neural networks (GNNs)}
GNN takes a graph $G=(\boldsymbol{V},\boldsymbol{E})$ as an input, computes properties and updates the graph, and outputs an updated graph $G'=(\boldsymbol{V}', \boldsymbol{E}')$  with an identical structure, where $\boldsymbol{V}'$ and $\boldsymbol{E}'$ are the set of updated vertex and edge features ($\boldsymbol{v}_i'$ and $\boldsymbol{e}_{i, j}'$). Message passing is the process of updating the graph by propagating information through it. 

\begin{figure}[!htbp]
    \centering
    \includegraphics[width=1.0\textwidth]{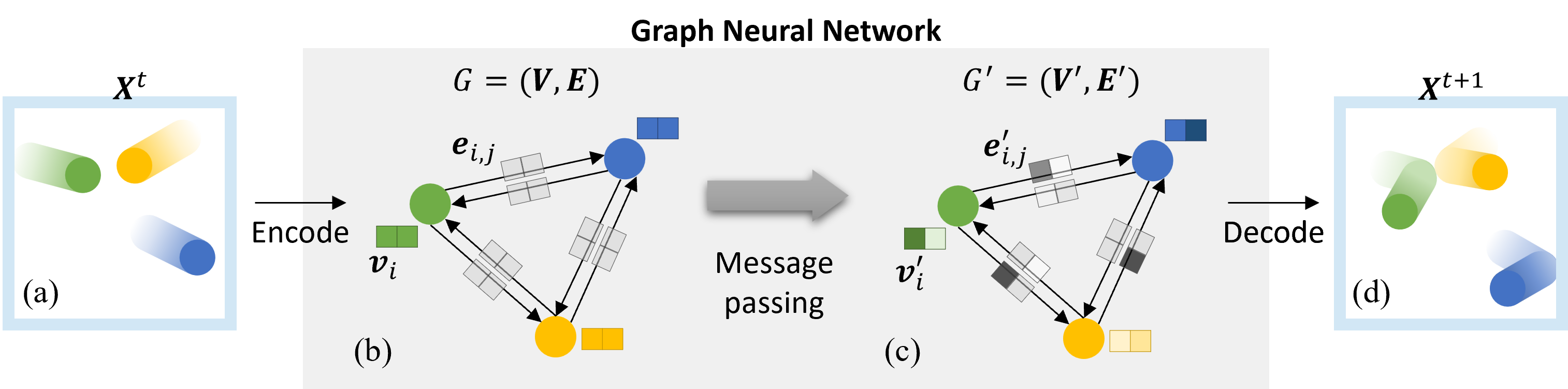}
    \caption{An example of a graph and graph neural network (GNN) that process the graph (modified from \cite{Battaglia2018}): (a) A state of the current physical system ($\boldsymbol{X}_t$) where the grains are bouncing in a box boundary; (b) Graph representation of the physical system ($G$). There are three vertices representing grains and six edges representing their directional interaction shown as arrows; (c) The updated graph ($G'$) that GNN outputs through message passing; (d) The predicted future state of the physical system ($\boldsymbol{X}_{t+1}$) (i.e., the positions of the grains at the next timestep) decoded from the updated graph.}
    \label{fig:balls-in-box}
\end{figure}

In the grains-in-a-box example, the GNN first takes the original graph $G=(\boldsymbol{V}, \boldsymbol{E})$  (\cref{fig:balls-in-box}b) that describes the current state of the physical system ($\boldsymbol{X}_t$). The GNN then updates the state of the physical system through message passing, which models the exchange of energy and momentum between the grains, and returns an updated graph $G'=(\boldsymbol{V}',\boldsymbol{E}')$ (\cref{fig:balls-in-box}c). We decode $G'$, the output of GNN, to extract information related to the future state of the physical system ($\boldsymbol{X}_{t+1}$), such as the next position or acceleration of the grains (\cref{fig:balls-in-box}d). 

\subsubsection{Message passing} \label{message_passing}
Message passing consists of three operations: message construction (\cref{eq:message-construct}), message aggregation (\cref{eq:message-aggregate}), and the vertex update function (\cref{eq:node-update}).

\begin{equation}\label{eq:message-construct}
\boldsymbol{e}_{i,j}^\prime=\phi_{\boldsymbol{\Theta}_\phi}\left(\boldsymbol{v}_i,\boldsymbol{v}_j,\boldsymbol{e}_{i,j}\right)
\end{equation}

\begin{equation}\label{eq:message-aggregate}
{\bar{\boldsymbol{v}}}_i=\sum_{j\in N\left(i\right)} \boldsymbol{e}_{i,j}^\prime
\end{equation}

\begin{equation}\label{eq:node-update}
\boldsymbol{v}_i^\prime=\gamma_{\boldsymbol{\Theta}_\gamma}\left(\boldsymbol{v}_i,{\bar{\boldsymbol{v}}}_i\right)
\end{equation}

\begin{figure}[!htbp]
    \centering
    \includegraphics[width=1.0\textwidth]{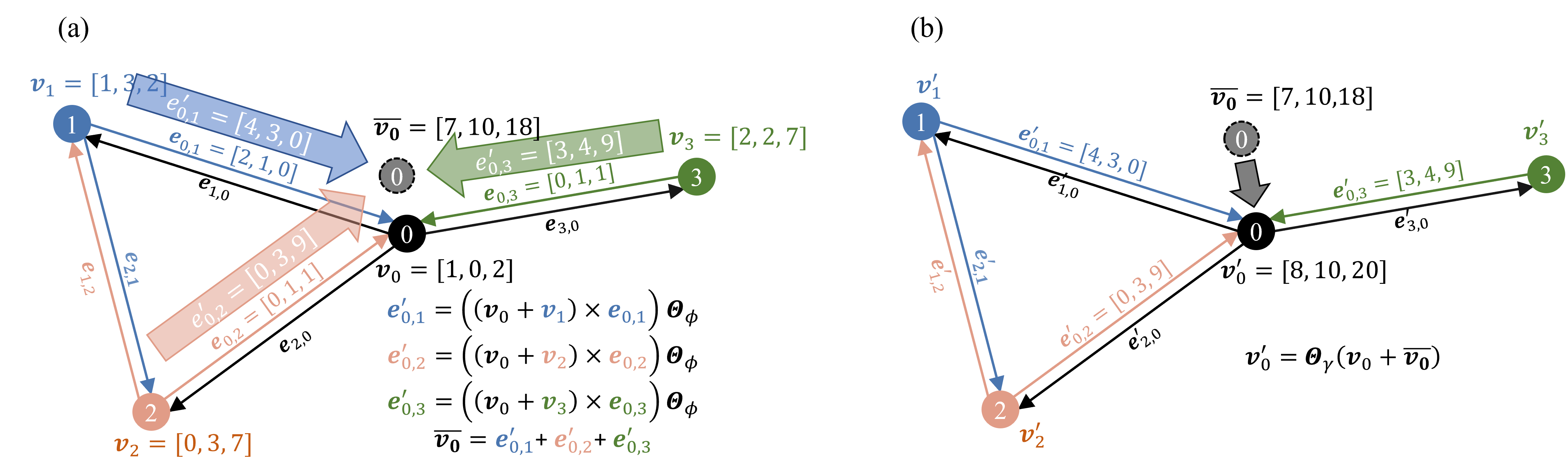}
    \caption{An example of message passing on a graph: (a) message construction directing to receiver vertex 0 ($\boldsymbol{e}_{0,\ 1}^\prime$,  $\boldsymbol{e}_{0,\ 2}^\prime$,  $\boldsymbol{e}_{0,\ 3}^\prime$) and the resultant aggregated message ($\boldsymbol{{\bar{v}}}_0$); (b) feature update at vertex 0 using $\boldsymbol{{\bar{v}}}_0$. Note that we assume $\boldsymbol{\Theta}_\phi$ and $\boldsymbol{\Theta}_r$ are 1.0 for the convenience of calculation.}
    \label{fig:message-passing}
\end{figure}

The subscript $\boldsymbol{\Theta}_\phi$ and $\boldsymbol{\Theta}_\gamma$ represent a set of learnable parameters in each computation. The message construction function $\phi_{\boldsymbol{\Theta}_\phi}$ (\cref{eq:message-construct}) takes the feature vectors of the receiver and sender vertices ($\boldsymbol{v}_i$ and $\boldsymbol{v}_j$) and the feature vector of the edge connecting them ($\boldsymbol{e}_{i,\ j}$) and returns an updated edge feature vector $\boldsymbol{e}_{i,j}^\prime$ as the output. $\phi_{\boldsymbol{\Theta}_\phi}$ is a matrix operation including the learnable parameter $\boldsymbol{\Theta}_\phi$. The updated edge feature vector $\boldsymbol{e}_{i,j}^\prime$ is the message sent from vertex $j$ to $i$. \Cref{fig:message-passing}a shows an example of constructing messages on edges directed to vertex 0 originating from vertices 1, 2, and 3 ($\boldsymbol{e}_{0,\ 1}^\prime$, $\boldsymbol{e}_{0,\ 2}^\prime$, $\boldsymbol{e}_{0,\ 3}^\prime$). Here, we define the message construction function $\phi_{\boldsymbol{\Theta}_\phi}$ as $\left(\left(\boldsymbol{v}_i+\boldsymbol{v}_j\right)\times \boldsymbol{e}_{i,\ j}\right)\times\boldsymbol{\Theta}_\phi$. The updated feature vector $\boldsymbol{e}_{0,\ 1}^\prime$ is computed as $\left(\left(\boldsymbol{v}_0+\boldsymbol{v}_1\right)\times \boldsymbol{e}_{0,\ 1}\right)\times\boldsymbol{\Theta}_\phi$, where $\boldsymbol{v}_0$ and $\boldsymbol{v}_1$ are the receiver and sender vertex feature vectors, and $\boldsymbol{e}_{0, \ 1}$ is their edge feature vector. Suppose we assume all values of $\boldsymbol{\Theta}_\phi$ are 1.0 for simplicity, we obtain $\boldsymbol{e}_{0,\ 1}^\prime=(\left(\left[1,\ 0,\ 2\right]\right)+\left[1,\ 3,\ 2\right])\times\left[2,\ 1,\ 0\right]^T)\times1=[4,\ 3,\ 0]$. Similarly, we can compute the messages $\boldsymbol{e}_{0,\ 2}^\prime=\left[0,\ 3,\ 9\right]$ and $\boldsymbol{e}_{0,\ 3}^\prime=\left[3,\ 4,\ 9\right]$. 

The next step in message passing is the message aggregation $\mathrm{\Sigma}_{j\in N\left(i\right)}$ (\cref{eq:message-aggregate}), where $N(i)$ is the set of sender vertices $j$ related to vertex $i$). It collects all the messages directing to vertex $i$ and aggregates those into a single vector with the same dimension as the aggregated message ($\boldsymbol{\bar{v}}_i$). The aggregation rule can be element-wise vector summation or averaging; hence it is a permutation invariant computation. In \cref{fig:message-passing}a, the aggregated message $\boldsymbol{\bar{v}}_0=[7,\ 10,\ 18]$ is the element-wise summation of the messages directing to vertex 0 as $\boldsymbol{\bar{v}}_0=\boldsymbol{e}_{0,\ 1}^\prime+\ \boldsymbol{e}_{0,\ 2}^\prime+\ \boldsymbol{e}_{0,\ 3}^\prime$. 

The final step of the message passing is updating vertex features using \cref{eq:node-update}. It takes the aggregated message ($\boldsymbol{\bar{v_i}}$) and the current vertex feature vector $\boldsymbol{v}_i$, and returns an updated vertex feature vector $\boldsymbol{v}_i^\prime$, using predefined vector operations including the learnable parameter $\boldsymbol{{\Theta}}_\gamma$. \Cref{fig:message-passing}b shows an example of the update at vertex 0. Here, we define the update function $\gamma_{\boldsymbol{\Theta}_\gamma}$ as $\boldsymbol{\Theta}_\gamma\left(\boldsymbol{v}_i+{\boldsymbol{\bar{v}}}_i\right)$. The updated  feature vector $\boldsymbol{v}_0^\prime$ is computed as $\boldsymbol{\Theta}_\gamma\left(\boldsymbol{v}_0+{\boldsymbol{\bar{v}}}_0\right)$. Assuming all parameters in $\boldsymbol{\Theta}_\gamma$ are 1.0 for simplicity, we obtain $\boldsymbol{v}_0^\prime= 
\boldsymbol{v}_0 + \bar{\boldsymbol{v}}_0 =  \left[1,\ 0,\ 2\right]+\left[7,\ 10,\ 18\right]=[8,\ 10,\ 20]$. Similarly, we update the other vertex features $(\boldsymbol{v}_1^\prime, \ \boldsymbol{v}_2^\prime, \ \boldsymbol{v}_3^\prime)$.

After message passing, the graph vertex and edge features ($\boldsymbol{v}_i$ and $\boldsymbol{e}_{i,\ j}$) are updated to $\boldsymbol{v}_i^\prime$ and $\boldsymbol{e}_{i,\ j}^\prime$. The GNN may include multiple message passing steps to propagate the information further through the network.

Unlike the example shown above, where we assume a constant value of 1.0 for the learnable parameters, in a supervised learning environment, the optimization algorithm will find a set of the best learnable parameters ($\boldsymbol{\Theta}_\phi, \boldsymbol{\Theta}_\gamma$) in the message passing operation.

\subsection{Graph Neural Network-based Simulator (GNS)}
\begin{figure}[!htbp]
    \centering
    \includegraphics[width=1.0\textwidth]{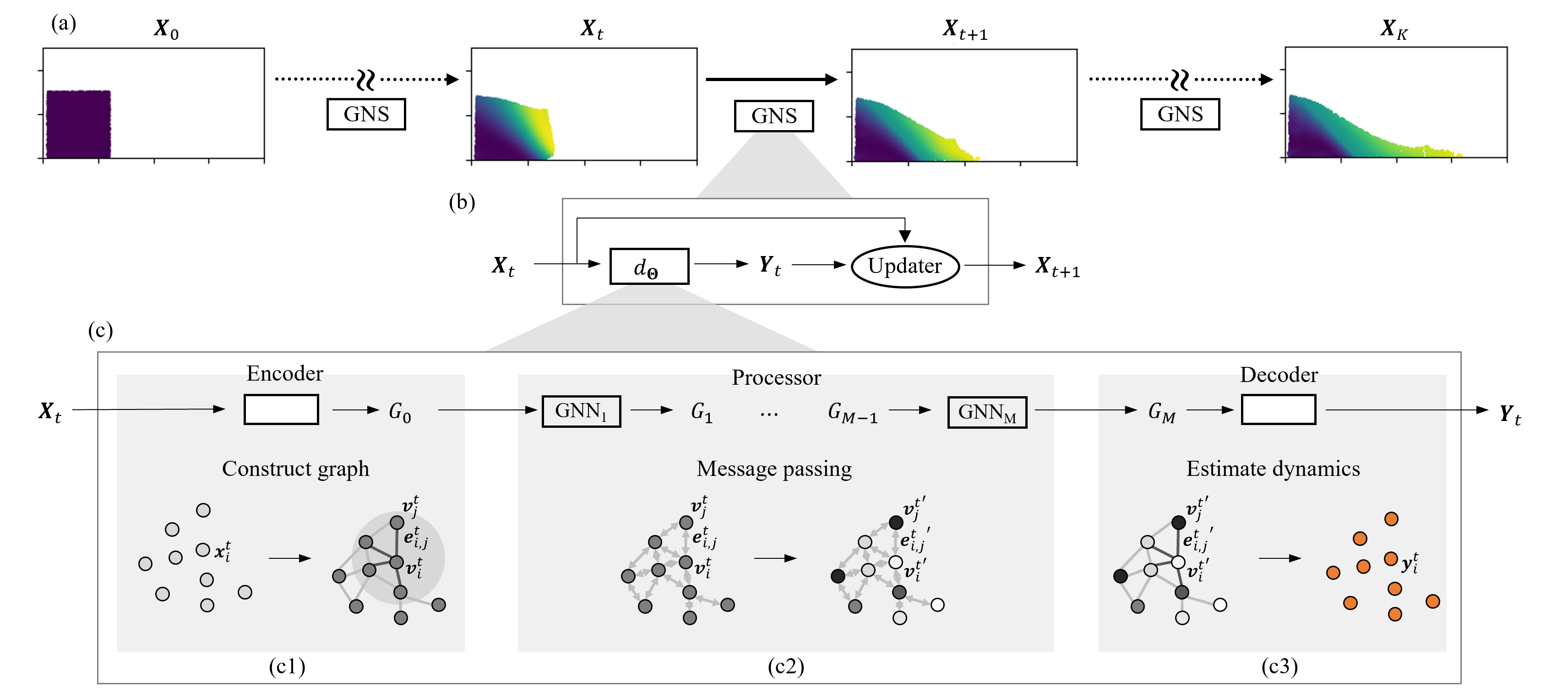}
    \caption{The structure of the graph neural network (GNN)-based physics simulator (GNS) for granular flow (modified from \cite{Sanchez2020}): (a) The entire simulation procedure using the GNS, (b) The computation procedure of GNS and its composition, (c) The computation procedure of the parameterized function approximator $d_\mathrm{\Theta}$ and its composition.}
    \label{fig:gns}
\end{figure}

In this study, we use GNN as a surrogate simulator to model granular flow behavior. \Cref{fig:gns} shows an overview of the general concepts and structure of the GNN-based simulator (GNS) proposed by \cite{Sanchez2020}. Consider a granular flow domain represented as material points (\cref{fig:gns}a), which represent the collection of grains. In GNS, we represent the physical state of the granular domain at time $t$ with a set of $\boldsymbol{x}_i^t$ describing the state and properties of each material point. The GNS takes the current state of the granular flow $\boldsymbol{x}_t^i\in \boldsymbol{X}_t$ and predicts its next state ${\boldsymbol{x}_{i+1}^i\in \boldsymbol{X}}_{t+1}$ (\cref{fig:gns}a). The GNS consists of two components: a parameterized function approximator $d_\mathrm{\Theta}$ and an updater function (\cref{fig:gns}b). The function approximator $d_\mathrm{\Theta}$ takes $\boldsymbol{X}_t$ as an input and outputs dynamics information ${\boldsymbol{y}_i^t\in \boldsymbol{Y}}_t$. The updater then computes $\boldsymbol{X}_{t+1}$ using $\boldsymbol{Y}_t$ and $\boldsymbol{X}_t$. \Cref{fig:gns}c shows the details of $d_\mathrm{\Theta}$ which consists of an encoder, a processor, and a decoder. The encoder (\cref{fig:gns}-c1) takes the state of the system $\boldsymbol{X}^t$ and embeds it into a latent graph $G_0=\left(\boldsymbol{V}_0,\ \boldsymbol{E}_0\right)$ to represent the relationships between material points. The vertices $\boldsymbol{v}_i^t\in \boldsymbol{V}_0$ contain latent information of the current state of the material point, and the edges $\boldsymbol{e}_{i,j}^t\in \boldsymbol{E}_0$ contain latent information of the pair-wise relationship between material points. Next, the processer (\cref{fig:gns}-c2) converts the input graph $G_0$ to the output graphs $G_M$ through $M$ stacks of message-passing GNN ($G_0\rightarrow G_1\rightarrow \cdots \rightarrow G_M$). The message passing computes the interaction between vertices. Finally, the decoder (\cref{fig:gns}-c3) extracts the dynamics of the points ($\boldsymbol{Y}^t$) from $G_M$, such as the acceleration of the physical system. The entire simulation (\cref{fig:gns}a) involves running GNS surrogate model through $K$ timesteps predicting from the initial state $\boldsymbol{X}_0$ to $\boldsymbol{X}_K \left(\boldsymbol{X}_0, \ \boldsymbol{X}_1, \ \ldots, \ \boldsymbol{X}_K\right)$, updating at each step ($\boldsymbol{X}_t\rightarrow \boldsymbol{X}_{t+1}$). We call this successive prediction from GNS the "rollout".

In the following sections, we explain the details of our input $\boldsymbol{X}^t$ (\cref{fig:gns}a), the encoder, processor, and decoder in $d_\mathrm{\Theta}$ (\cref{fig:gns}c), and how we compute $\boldsymbol{X}^{t+1}$ from $\boldsymbol{X}^t$ using the GNS updater function (\cref{fig:gns}b).

\subsubsection{Input}
The input to the GNS, $\boldsymbol{x}_i^t\in \boldsymbol{X}^t$ (\cref{eq:inputs}), is a vector consisting of the current material point position $\boldsymbol{p}_i^t$, the material point velocity context ${\boldsymbol{\dot{p}}}_i^{\le t}$, information on boundaries $\boldsymbol{b}_i^t$, and material point type embedding $\boldsymbol{f}$. The current state $\boldsymbol{x}_i^t$ will be used to construct vertex feature ($\boldsymbol{v}_i^t$) (\cref{eq:encode}).

\begin{equation}\label{eq:inputs}
\boldsymbol{x}_i^t=[\boldsymbol{p}_i^t,\ {\boldsymbol{\dot{p}}}_i^{\le t},\ \boldsymbol{b}_i^t,\ \boldsymbol{f}, \ \boldsymbol{m}]
\end{equation}

The velocity context $\boldsymbol{{\dot{p}}}_i^{\le t}$ includes the current and previous material point velocities for $n$ timesteps $\left[\boldsymbol{{\dot{p}}}_i^{t-n},\cdots,\ {\boldsymbol{\dot{p}}}_i^t\right]$ with $n+1$ velocities. We use $n=4$ to include sufficient velocity context in the vertex feature $\boldsymbol{v}_i^t$. \cite{Sanchez2020} show that having $n>1$ significantly improves the model performance. We compute the velocities using the finite difference of the position sequence (i.e., ${\boldsymbol{\dot{p}}}_i^t=\left(\boldsymbol{p}_i^t-\boldsymbol{p}_i^{t-1}\right)/\mathrm{\Delta}t$). $\boldsymbol{b}_i^t$ is boundary information. For a 2D problem, $\boldsymbol{b}_i^t$ has four components, each indicating the distance between material points and the four walls. We normalize $\boldsymbol{b}_i^t$ by the connectivity radius $R$ which defines the interaction zone, explained in the next section, and restrict it between -1.0 to 1.0. $\boldsymbol{b}_i^t$ is used to evaluate the role of boundary friction on material points. $\boldsymbol{f}$ is a vector embedding distinguishing material point types such as deformable or rigid material.  $\boldsymbol{m}$ is an optional feature that represents the material properties of the corresponding material points. If the training data includes different material properties for each material point, we can simply add relevant features here.

\added{The GNS model accounts for the friction between vertices representing material points and boundaries by monitoring their distance using $\boldsymbol{b}_i^t$ as outlined in~\cref{eq:inputs}. GNS predicts the frictional acceleration when the nodal boundary distance is less than the connectivity radius $R$. When the boundary distance is larger than $R$, we clip the $\boldsymbol{b}_i^t$ to -1 or 1, signaling that boundary friction does not directly affect the material points at this distance. During training, GNS learns the interaction between the material points' boundary distance and frictional acceleration from training data. The MPM training data models the tangential friction force as a function of the normal acceleration and the friction coefficient. With the mass of the material points known, we can compute the frictional forces from the accelerations predicted near the boundary.}

We define the interaction between material points $i$ and $j$ as $\boldsymbol{r}_{i,\ j}^t$ using the distance and displacement of the material points in the current timestep (see \cref{eq:interaction}). The former reflects the level of interaction, and the latter reflects its spatial direction. $\boldsymbol{r}_{i,\ j}^t$ will be used to construct edge features ($\boldsymbol{e}_{i,j}^t$). Although the distance provides an additional physical context about the interactions, we found that omitting it from \cref{eq:interaction} does not necessarily harm the model's performance since the distance can be calculated with the displacement. Specifically, the model with the distance feature shows the mean squared error of 0.0016 $m$ to 0.0018 $m$, whereas the model without it shows 0.0014 $m$ to 0.0017 $m$, when we validate the model's performance in predicting the evolution of material point positions.

\begin{equation}\label{eq:interaction}
\boldsymbol{r}_{i,j}^t=\left[\left(\boldsymbol{p}_i^t-\boldsymbol{p}_j^t\right), \ \lVert \boldsymbol{p}_i^t-\boldsymbol{p}_j^t \rVert\right]
\end{equation}

\subsubsection{Encoder}
The vertex and edge encoders ($\varepsilon_{\boldsymbol{\Theta}}^{v}$ and $\varepsilon_{\boldsymbol{\Theta}}^{e}$) convert $\boldsymbol{x}_i^t$ and $\boldsymbol{r}_{i,\ j}^t$ into the vertex and edge feature vectors ($\boldsymbol{v}_i^t$ and $\boldsymbol{e}_{i,j}^t$) (\cref{eq:encode}) and embed them into a latent graph $G_0=(\boldsymbol{V}_0,\ \boldsymbol{E}_0)$,  $\boldsymbol{v}_i^t\in\ \boldsymbol{V}_0$, $\boldsymbol{e}_{i,j}^t\in\ \boldsymbol{E}_0$. 

\begin{equation}\label{eq:encode}
\boldsymbol{v}_i^t=\varepsilon_{\boldsymbol{\Theta}}^{v} \left(\boldsymbol{x}_i^t\right), \\\ \boldsymbol{e}_{i,j}^t=\varepsilon_{\boldsymbol{\Theta}}^{e}\left(\boldsymbol{r}_{i,j}^t\right)
\end{equation}

We use a two-layered 128-dimensional multi-layer perceptron (MLP) for the $\varepsilon_{\boldsymbol{\Theta}}^{v}$ and $\varepsilon_{\boldsymbol{\Theta}}^{e}$. The MLP and optimization algorithm search for the best candidate for the parameter set $\boldsymbol{\Theta}$ that estimates a proper way of representing the physical state of the material points and their relationship which will be embedded into $G_0$. 

The edge encoder $\varepsilon_{\boldsymbol{\Theta}}^v$ uses $\boldsymbol{x}_i^t$ (\cref{eq:inputs}) without the current position of the material point ($\boldsymbol{p}_i^t$), but with its velocities ($\boldsymbol{{\dot{p}}}_i^{\le t}$), as velocity governs the momentum, and the interaction dynamics is independent of the absolute position of the material points. \cite{rubanova2021constraint} confirmed that including position causes poorer model performance. We only use $\boldsymbol{p}_i^t$ to predict the next position  $\boldsymbol{p}_i^{t+1}$ based on the predicted velocity $\boldsymbol{{\dot{p}}}_i^{t+1}$ using Explicit Euler integration. 

We consider the interaction between two material points by constructing edges between them all pairs of vertices located within a certain distance called connectivity radius $R$ (see the shaded circular area in \cref{fig:gns}b). The connectivity radius is a critical hyperparameter that governs how effectively the model learns the local interaction. $R$ should be sufficiently large to include the local interaction between material points and capture the simulation domain's global dynamics.  

\subsubsection{Processor}
The processor performs message passing (based on \cref{eq:message-construct} to \cref{eq:node-update}) on the initial latent graph ($G_0$) from the encoder for $M$ times ($G_0\rightarrow G_1\rightarrow \cdots \rightarrow G_M$) and returns a final updated graph $G_M$. We use two-layered 128-dimensional MLPs for both the message construction function $\phi_{\boldsymbol{\Theta}_\phi}$ and vertex update function $\gamma_{\boldsymbol{\Theta}_r}$, and element-wise summation for the message aggregation function $\mathrm{\Sigma}_{j\in N\left(i\right)}$ in \cref{eq:message-construct} to \cref{eq:node-update}. We set $M=10$ to ensure sufficient message propagation through the network. These stacks of message passing model information propagation through the network of material points.

\subsubsection{Decoder}
The decoder $\delta_{\boldsymbol{\Theta}}^v$ extracts the dynamics $\boldsymbol{y}_i^t\in \boldsymbol{Y}^t$ of the material points from the vertices ${\boldsymbol{v}_i^t}^\prime$ (\cref{eq:decode}) using the final graph $G_M$. We use a two-layered 128-dimensional MLP for $\delta_{\boldsymbol{\Theta}}^v$, which learns to extract the relevant dynamics for material points from $G_M$.

\begin{equation}\label{eq:decode}
\boldsymbol{y}_i^t=\delta_{\boldsymbol{\Theta}}^v\left({\boldsymbol{v}_i^t}^\prime\right)
\end{equation}

\subsubsection{Updater}\label{sec:updater}
We use the dynamics $\boldsymbol{y}_i^t$ to predict the velocity and position of the material points at the next timestep (${\boldsymbol{\dot{p}}}_i^{t+1}$ and  $\boldsymbol{p}_i^{t+1}$) based on Euler integration (\cref{eq:inertial} and \cref{eq:static}), which makes $\boldsymbol{y}_i^t$ analogous to acceleration $\boldsymbol{{\ddot{p}}}_i^t$.

\begin{equation}\label{eq:inertial}
\boldsymbol{{\dot{p}}}_i^{t+1}=\boldsymbol{\dot{p}}_i^t+\boldsymbol{y}_i^t\mathrm{\Delta t}
\end{equation}

\begin{equation}\label{eq:static}
\boldsymbol{p}_i^{t+1}=\boldsymbol{p}_i^t+\boldsymbol{\dot{p}}_i^{t+1}\mathrm{\Delta t}
\end{equation}

Based on the new position and velocity of the material points, we update $\boldsymbol{x}_i^t\in \boldsymbol{X}^t$ (\cref{eq:inputs}) to $\boldsymbol{x}_i^{t+1}\in \boldsymbol{X}^{t+1}$. The updated physical state $\boldsymbol{X}^{t+1}$ is then used to predict the position and velocity for the next timestep.

The updater imposes inductive biases, such as an inertial frame, on GNS to force it only to learn the interaction dynamics, improving learning efficiency. A traditional neural network learns both the update scheme and the interaction dynamics:
\begin{equation}
    p^{t+1} = NN(p^t, v^t) \,.
\end{equation}
Whereas, using an inertial prior, we force the GNS only to learn the interaction dynamics, by hardcoding the update function:
\begin{equation}
    p^{t+1} = p^t + v^t \cdot \Delta t + NN(p^t, v^t) \,.
\end{equation}
Nevertheless, GNS does not directly predict the next position from the current position and velocity (i.e., $\boldsymbol{p}_i^{t+1}=GNS\left(\boldsymbol{p}_i^t,\ \boldsymbol{\dot{p}}_i^t\right)$) which has to learn the static motion and inertial motion. Instead, it uses (1) the inertial prior (\cref{eq:inertial}) where the prediction of the next velocity $\boldsymbol{\dot{p}}_i^{t+1}$ should be based on the current velocity $\boldsymbol{\dot{p}}_i^t$  and (2) the static prior (\cref{eq:static}) where the prediction of the next position $\boldsymbol{p}_i^{t+1}$ should be based on the current position $\boldsymbol{p}_i^t$. These make GNS focus on learning unknown dynamics by hardcoding known physics. Since GNS learns the dynamics of material points through interactions independent of absolute position, GNS is generalizable to other geometric conditions.

\section{Training and evaluation}

We now train the GNS to predict granular column collapse and granular flow interacting with barriers. This section explains how we generate training data, details of the training process, and how we evaluate the performance of the GNS.

\subsection{Material Point Method (MPM)}
We utilize the Material Point Method (MPM) to generate the GNS training dataset of granular flow simulations. The MPM is a hybrid Eulerian-Lagrangian approach designed for modeling large-deformation flows \citep{soga2016}. In the MPM, a continuum body is represented by individual material points that traverse a static background grid. The governing equation is solved at the nodes, and the updated velocity field is subsequently mapped back to the material points. We employ the position information stored in these material points to construct the current state $\boldsymbol{X}^t$ in the GNS. For more information on MPM refer to~\cite {soga2016}. 

\subsection{Datasets}\label{sec:datasets}
\subsubsection{Granular column collapse}
We prepare three types of datasets (\cref{table:train_data}): granular column collapse with a single material property, granular column collapse with multiple material properties, and granular flow interacting with barriers in three-dimensional space. 

For the granular column collapse case, the training datasets include 26 granular flow trajectories of square-shaped granular mass in a two-dimensional box boundary simulated by the MPM explicit time integration method using the CB-Geo MPM code \citep{kumar2019scalable}. Each simulation has a different initial configuration regarding the size of the square granular mass, position, and velocity. \Cref{table:train_data} presents the details of the simulation configurations. The datasets are published on DesignSafe \citep{choi2023_data}. \ref{appendix:training_data} shows all the training trajectories with different initial configurations and initial velocities.

We also create the validation datasets to check if the model experiences overfitting. The datasets include seven trajectories of randomly picked rectangular-shaped granular mass with different initial configurations not included in the training datasets. 

Additionally, we prepare another dataset that involves granular column collapse with multiple friction angles ($\phi$) to evaluate the ability of GNS to learn diverging granular flow behaviors depending on the friction angle of the material. It includes MPM simulations with five different friction angles ($\phi = 15\degree, \ 22.5\degree, \ 30\degree, \ 37.5\degree, \ 45\degree$) where each case has 60 simulations. To include the material-dependent characteristic in GNS, we simply use the normalized friction angle ($\tan{\phi}$) as $\boldsymbol{m}$ in \cref{eq:inputs} for the additional vertex feature. Similar to the dataset with a single friction angle, the initial geometry of the granular mass is restricted to the square shape.

\subsubsection{Granular flow with barriers}
For the barrier interaction case, the training datasets include 890 granular flow trajectories of granular mass interaction with barriers in a three-dimensional box boundary. We use Taichi MPM code \citep{hu2019taichi} with explicit time integration to generate the simulation data. The initial geometry of the granular mass is limited to a cube shape with varying lengths and initial velocities. For each simulation, the granular material interacts with one or two $0.1\times0.3\times0.1 \ m$ (width, height, and length) shaped barriers in $0.8\times0.8\times0.8 m$ domain. The barriers are modeled as a rigid body. The details of the simulation configurations are presented in \Cref{table:train_data}.

\begin{table}[]
\centering
\centering
\caption{Details of the Material Point Method (MPM) simulation geometries and properties used for generating the training datasets.}
\label{table:train_data}
\resizebox{\textwidth}{!}{%
\begin{tabular}{lllll} 
\toprule
\multicolumn{2}{c}{\multirow{2}{*}{Property}} & \multicolumn{3}{c}{Dataset} \\ 
\cmidrule{3-5}
\multicolumn{2}{c}{} & \multicolumn{1}{c}{Granular column collapse} & \begin{tabular}[c]{@{}l@{}}Multi-material \\granular column collapse\end{tabular} & \begin{tabular}[c]{@{}l@{}}Granular flow \\with barriers\end{tabular} \\ 
\midrule
\multicolumn{2}{l}{Simulation boundary} & 1.0×1.0~$m$ & 1.0×1.0 $m$ & 0.8×0.8×0.8~$m$ \\
\multicolumn{2}{l}{MPM element length} & 0.025×0.025 $m$ & 0.01×0.01 $m$ & 0.03125×0.03125×0.03125 $m$ \\
\multicolumn{2}{l}{Material point configuration} & 25,600 $points/m^2$ & 40,000 $points/m^2$ & 262,144 $points/m^3$ \\
\multicolumn{2}{l}{\vcell{Granular mass geometry}} & \vcell{0.2×0.2 and~0.3×0.3~$m$} & \vcell{0.2×0.2 to 0.4×0.4 $m$} & \vcell{0.2×0.2×0.2 to~0.4×0.4×0.4 $m$} \\[-\rowheight]
\multicolumn{2}{l}{\printcelltop} & \printcellmiddle & \printcellmiddle & \printcellmiddle \\
\multicolumn{2}{l}{Max. number of particles} & 2.3K & 6.4K & 17K \\
\multicolumn{2}{l}{Barrier geometry} & None & None & 0.1×0.3×0.1 $m$ \\
\multicolumn{2}{l}{Simulation duration~(timesteps)} & 400~(dt=0.0025 s) & 400~(dt=0.0025 s) & 350 (dt=0.0025 s) \\ 
\midrule
\multirow{7}{*}{Material property} & Model & Mohr-Coulomb & Mohr-Coulomb & Mohr-Coulomb \\
 & Density & 1,800 $kg/m^3$ & 1,800 $kg/m^3$ & 1,800 $kg/m^3$ \\
 & Youngs modulus & 2 $GPa$ & 2 $GPa$ & 2 $GPa$ \\
 & Poisson ratio & 0.3 & 0.3 & 0.3 \\
 & Friction angle & 30$\degree$ & 15, 17.5, 22.5, 30, 37.5, 45 $\degree$ & 35$\degree$ \\
 & Cohesion & 0.1 $kPa$ & 0.1 $kPa$ & None \\
 & Tension cutoff & 0.05 $kPa$ & 0.05 $kPa$ & None \\
\bottomrule
\end{tabular}
}
\end{table}

\subsection{Training}
Our GNS has a learnable parameter set $\mathrm{\Theta}$. We train $\mathrm{\Theta}$ to minimize the loss calculated as the mean squared error (MSE) between $\boldsymbol{y}_i^t$ (predicted proxy-acceleration) and the ground truth acceleration $\boldsymbol{\ddot{p}}_i^t$ for all material points $i=1,\ 2,\ \ldots,\ N$ as shown in \cref{eq:objective} based on gradient ($\mathrm{\nabla} loss_\mathrm{\Theta}$)-based optimizer, Adam \citep{Kingma2014AdamAM}.

\begin{equation}\label{eq:objective}
    loss_\mathrm{\Theta}=\frac{1}{n}\sum_{i=1}^{N}\left(\boldsymbol{y}_i^t-\boldsymbol{\ddot{p}}_i^t\right)^2
\end{equation}

For training the GNS, we have to set hyperparameters to learn the flow behavior from the training trajectories properly. The first key hyperparameter is the connectivity radius $R$, which governs the model’s capacity to learn the interactions of material points. For each model that we train (granular column collapse and barrier interaction model), we employ different values of $R$ to consider a balance between the associated training costs and model capacity. For the granular column collapse model, we choose $R=0.030$ m, and for the barrier interaction model, we choose $R=0.025$, which includes about 9 to 10 and 3 to 4 material points along the diameter, respectively. The circular and spherical area defined by $R$ for each model incorporates approximately 70 and 17 material points inside. For the barrier interaction model, we design to have a smaller $R$, and correspondingly, fewer material points inside it, to avoid excessive edge formation, since the model is defined in the three-dimensional domain. Another important hyperparameter is the Gaussian noise value for perturbing the ground truth position in the training trajectories. Since every predicted position for each timestep is based on the previous prediction, which includes a prediction error, the simulation over the large timesteps is subjected to an exponential error accumulation. To avoid this issue, we train the model on input positions with Gaussian noise that emulates the prediction error made by a one-step prediction ($\boldsymbol{X}_t \rightarrow \boldsymbol{X}_{t+1}$). The inclusion of noise in training leads to more rigorous long-rollout predictions.

We use the learning rate ($lr$) decay with the initial value of $10^{-4}$ and decay rate of 0.1 ($lr=10^{-4} \times 0.1^{step/5\times 10^6}$) for more stable convergence. We use the batch size of two, i.e., $\boldsymbol{X}_t$ from two different trajectories are used simultaneously in updating the learnable parameters. We enable a multi-GPU training algorithm based on data-distributed parallelism. It distributes GNS models and datasets across GPUs, and the training is conducted in parallel for each GPU. In this way, the training process can observe as many datasets as the number of GPUs at each training step. For information on the scalability of the GNS algorithm, refer to \cite{kumar2022gns}.

\begin{figure}[!htbp]
    \centering
    \includegraphics[width=0.7\textwidth]{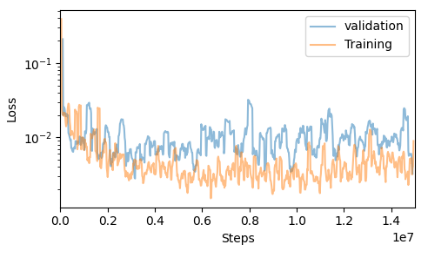}
    \caption{Evolution of GNS loss in training and validation with training steps.}
    \label{fig:loss_hist}
\end{figure}

For the granular column collapse model training, we investigate if the model experiences overfitting by plotting the loss history (\cref{fig:loss_hist}) for the training and validation datasets evaluated for every 10K training steps. The training and validation losses show a drastic decrease until 2M steps. After that, the validation loss tends to remain slightly larger than the training loss. \Cref{fig:loss_hist} shows no overfitting during the training.

\subsection{Granular column collapse}
\begin{figure}[!htbp]
    \centering
    \includegraphics[width=0.75\textwidth]{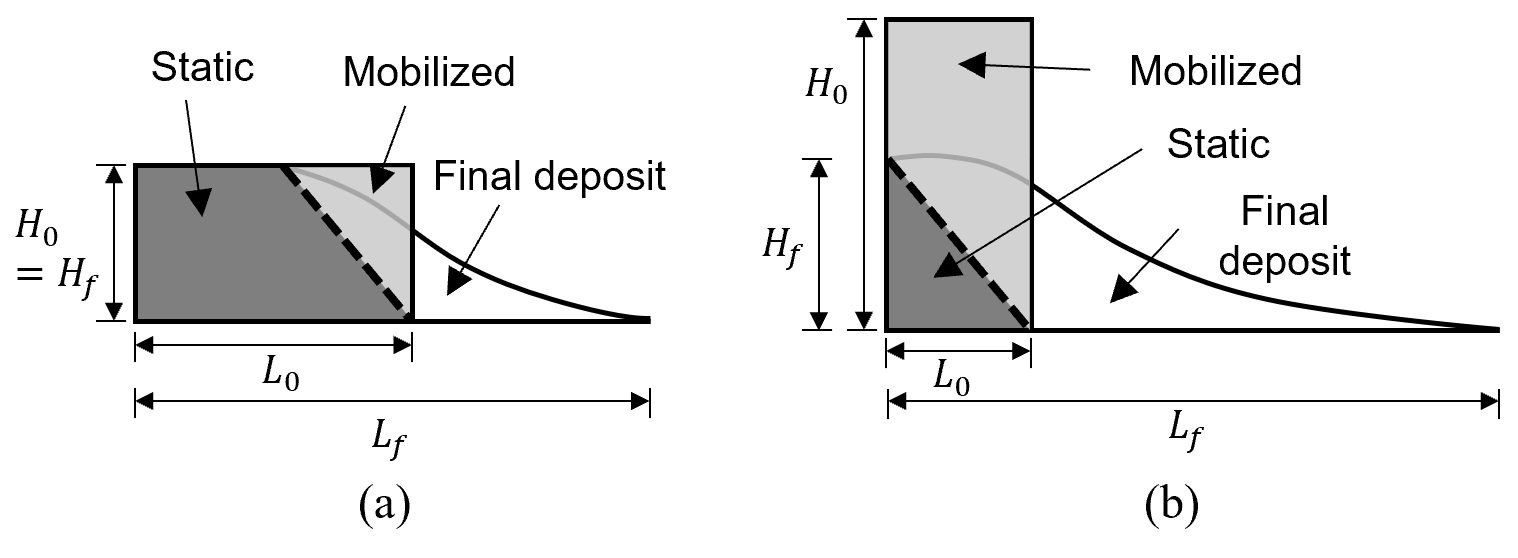}
    \caption{Schematics for the granular column collapse configuration and its behavior on the aspect ratio.}
    \label{fig:column_collapse}
\end{figure}

We trained the GNS to predict the collapse of a granular column (as studied by \cite{Lajeunesse2004, Lube2005}). \Cref{fig:column_collapse} shows the granular column collapse experiments to evaluate its ability to replicate granular flow dynamics. Granular column collapse is a simple physical experiment that captures the transient response of granular flow dynamics. The experiment involves the collapse of a granular column of initial height $H_0$ and length $L_0$ on a flat surface due to gravity. As the gate holding the column is removed, the granular material destabilizes, resulting in a runout. We measure the final runout deposit with the final height $H_f$ and runout $L_f$. 

The runout of the granular column is governed by the initial aspect ratio ($a=H_0/L_0$) \citep{staron2005, kumar2015thesis}. For short columns ($a\lesssim 2$) (\cref{fig:column_collapse}a), the soil mass fails along the flanks of the column above a well-defined failure surface (dashed line). The soil mass beneath the failure surface remains in static equilibrium throughout the collapse forming a truncated conical shape. With the increase in aspect ratio, the portion of the sliding mass above the failure surface increase, and the static part becomes smaller, forming a conical shape. For tall columns ($a\gtrsim2$) (\cref{fig:column_collapse}b), the majority of the soil mass is involved in the collapse, and it initially experiences a free fall due to gravitational acceleration. As the falling mass reaches the failure surface, the vertical kinetic energy is converted to horizontal acceleration, resulting in a longer runout distance than the short column (\cref{fig:column_collapse}a). 

In addition, researchers \citep{kumar2015thesis, staron2005, kermani2015, utili2015} observed a transition zone where the flow dynamics change from short to tall columns. The normalized runout ($\left(L_f-L_0\right)/L_0$) of a granular column is largely governed by the aspect ratio ($a$). The normalized runout represents how far the granular mass runs out before reaching the final deposit state compared to the initial length of the column. Short columns show a linear relationship with the initial aspect ratio. In contrast, tall columns have a power-law relationship with the initial aspect ratio. 

Another factor that affects the runout is the friction angle. \cite{nguyen2020friction} found that the increase in the friction angle tends to reduce runout since more shear stress is required to mobilize the granular mass. 

The GNS was trained only on the aspect ratio of 1.0. However, we evaluate its performance in predicting the runout dynamics of other aspect ratios by comparing the GNS predictions with the MPM simulations. \Cref{table:test_cases} presents the test cases for evaluating GNS performance trained on the single friction angle dataset.

\begin{table}[]
\centering
\caption{Granular column collapse simulation cases for testing GNS.}
\begin{tabular}{@{}cccccc@{}}
\toprule
\multicolumn{2}{c}{Test case}  & $H_0 \times L_0$ (m) & \begin{tabular}[c]{@{}c@{}}Duration \\ (timesteps)\end{tabular} & \begin{tabular}[c]{@{}c@{}}Simulation \\ boundary (m) \end{tabular}     & \begin{tabular}[c]{@{}c@{}}Number of \\ material points\end{tabular} \\ \midrule
\begin{tabular}[c]{@{}c@{}}Short \\ columns\end{tabular} & $a=0.5$ & 0.2 $\times$ 0.4 & 400  & \begin{tabular}[c]{@{}c@{}}X: 0 to 1.0\\ Y: 0 to 0.5\end{tabular}  & 1956  \\

 & $a=0.8$ & $0.24 \times 0.30$     & 400  & \begin{tabular}[c]{@{}c@{}}X: 0 to 1.0\\ Y: 0 to 0.5\end{tabular}  & 1824 \\

 & $a=1.0$ & 0.30 $\times$ 0.30     & 400 & \begin{tabular}[c]{@{}c@{}}X: 0 to 1.0\\ Y: 0 to 0.5\end{tabular}  & 2304  \\

\midrule

\begin{tabular}[c]{@{}c@{}}Tall \\ columns\end{tabular}  & $a=2.0$ & 0.30 $\times$ 0.15     & 400  & \begin{tabular}[c]{@{}c@{}}X: 0 to 1.0\\ Y: 0 to 0.5\end{tabular}  & 1152  \\

 & $a=3.0$ & 0.36 $\times$ 0.12     & 400  & \begin{tabular}[c]{@{}c@{}}X: 0 to 1.0\\ Y: 0 to 0.5\end{tabular}  & 1106  \\
 
 & $a=4.0$ & 0.35 $\times$ 0.075    & 400  & \begin{tabular}[c]{@{}c@{}}X: 0 to 1.0\\ Y: 0 to 0.5\end{tabular}  & 576 \\
 \midrule
 Up-scaled  & $a=0.8$ & 0.36 $\times$ 0.45     & 500  & \begin{tabular}[c]{@{}c@{}}X: 0 to 1.5 \\ Y: 0 to 1.0\end{tabular} & 5120 \\ \bottomrule
\end{tabular}
\label{table:test_cases}
\end{table}

\subsection{Granular flow interacting with barriers}
We test the GNS trained on granular flows interacting with barriers to predict its behaviors for debris-resisting baffles \citep{choi2014flume, choi2015computational, yang2020obstacle_numerical}. Debris-resisting baffles are rigid, flow-impeding structures installed perpendicular to potential landslide debris flow paths to mitigate excessive runout and related hazards downstream. 

Baffles work by slowing down the runout as it impacts each baffle, causing the flow to lose energy. As the debris diverges and passes through openings between baffles and hits the subsequent baffles, the energy of the flow dissipates. The energy dissipation caused by the baffles is quantified by the difference between the sum of kinetic and potential energy of the flow upstream and downstream of the baffles. It measures the general performance of the baffles for regulating energy from the flow.

Baffles can promote the deposition of material behind them and cause backwater effects upstream. The backwater effect refers to the increase in flow depth upstream of an obstruction like the baffles. A high rise of the upstream flow can cause hazardous overflow if it exceeds the baffle height, which can be uncontrollable. Therefore, upstream depth is an important criterion for baffle performance.

We evaluate the performance of GNS in terms of predicting overall runout dynamics, energy evolution, and upstream depth. Our training data only includes the cube-shaped granular mass geometry interacting with one or two barriers with the size of $0.1\times0.3\times0.1 \ m$ in the $0.8\times0.8\times0.8 \ m$ simulation domain. However, the prediction (i.e., flow with debris-resisting baffles) is performed with a cuboid-shaped granular mass colliding three barriers with the size of $0.15\times0.30\times0.15 \ m$ in the $1.8\times0.8\times1.8 \ m$ simulation domain, which is not seen during the training.

\section{Results and discussions}

In this section, we show the prediction performance of GNS on granular column collapse and granular flow interacting with barriers.

\subsection{Granular column collapse}

\begin{figure}[!htbp]
    \centering
    \includegraphics[width=0.70\textwidth]{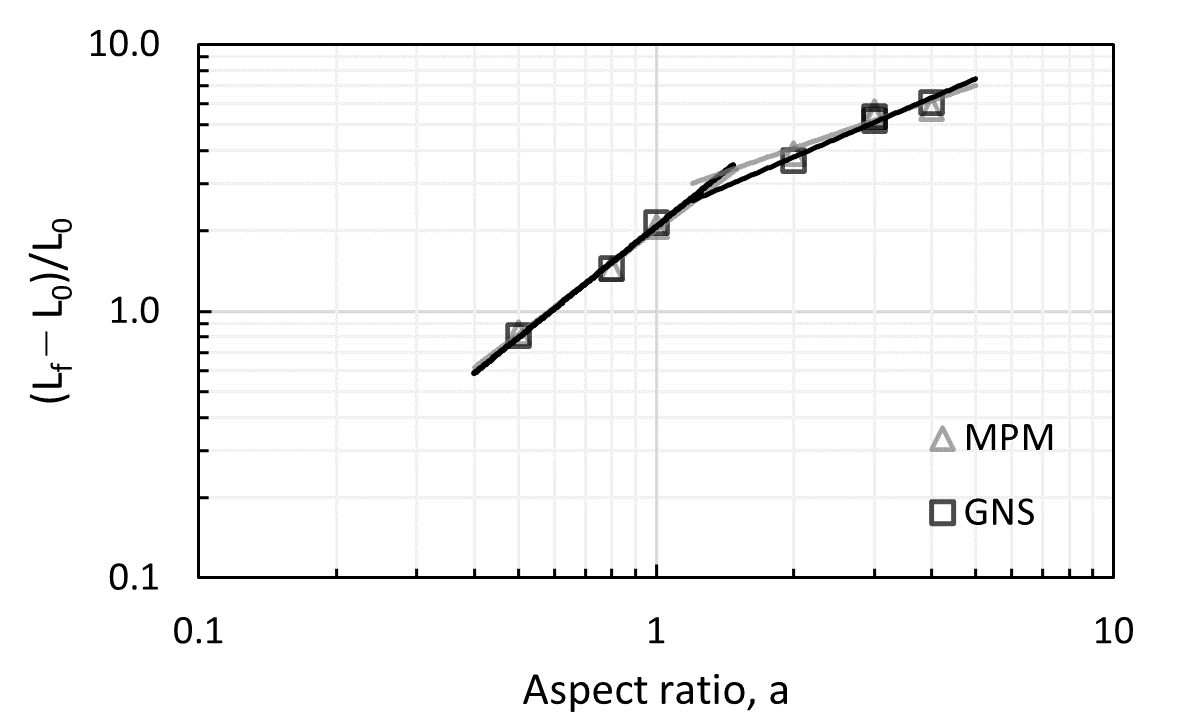}
    \caption{Normalized runout distance ($\left(L_f-L_0\right)/L_0$) with different aspect ratios ($a$).}
    \label{fig:powerlaw}
\end{figure}

We evaluate the GNS predictions of granular column collapse trained on a single friction described in \cref{table:train_data} against the MPM simulations in terms of the (1) geometry of sliding mass, (2) evolution of runout and height with time, and (3) energy evolution during the collapse. \Cref{fig:powerlaw} shows the normalized runout ($\left(L_f-L_0\right)/L_0$) predictions of GNS for different aspect ratios in comparison with MPM. $L_f$ is the distance from the left wall to the material point that runs out the farthest, as shown in \cref{fig:column_collapse}. Previous research observed a transition zone for the relationship between the normalized runout and aspect ratio that distinguishes short-column from tall-column dynamics. For both GNS and MPM, we observe the transition around an initial aspect ratio $a=1.2$ (\cref{fig:powerlaw}). \Cref{table:powerlaw} summarizes the errors between GNS predictions and MPM simulations for different aspect ratios. In general, the GNS runout prediction is within 5\% of the MPM runout estimate. \Cref{fig:powerlaw} suggests that the GNS successfully captures the dependence of the final runout with the initial aspect ratio, including the transition from the short to the tall column.

\begin{table}[]
\centering
\caption{Normalized runout from MPM and GNS depending on aspect ratios and corresponding prediction error.}
\begin{tabular}{@{}c >{\centering\arraybackslash}c >{\centering\arraybackslash}c c@{}}
\toprule
\multirow{2}{*}{Aspect ratio, $a$} & \multicolumn{2}{c}{Normalized runout} & \multirow{2}{*}{Runout error (\%)} \\ \cmidrule(lr){2-3}
& MPM  & GNS               &                                      \\ \midrule
0.5                                & 0.831             & 0.811             & 2.48                                 \\
0.8                                & 1.444             & 1.445             & 0.06                                 \\
1.0                                & 2.071             & 2.152             & 3.78                                 \\
2.0                                & 3.892             & 3.682             & 5.70                                 \\
3.0                                & 5.620             & 5.341             & 5.23                                 \\
4.0                                & 5.753             & 6.070             & 5.21                                 \\ \bottomrule
\end{tabular}
\label{table:powerlaw}
\end{table}

\subsubsection{Short column}
\begin{figure}[!htbp]
     \centering
     \begin{subfigure}[b]{0.8\textwidth}
         \centering
         \includegraphics[width=\textwidth]{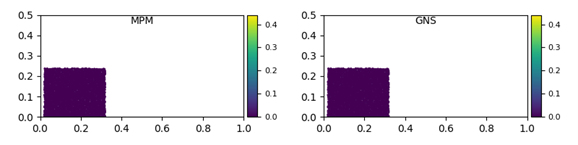}
         \caption{$t/\tau_c=0$}
         \label{fig:short1}
     \end{subfigure}
     \vfill
     \begin{subfigure}[b]{0.8\textwidth}
         \centering
         \includegraphics[width=\textwidth]{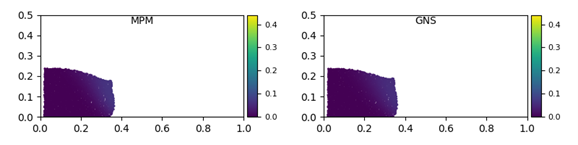}
         \caption{$t/\tau_c=1.0$}
         \label{fig:short2}
     \end{subfigure}
     \vfill
     \begin{subfigure}[b]{0.8\textwidth}
         \centering
         \includegraphics[width=\textwidth]{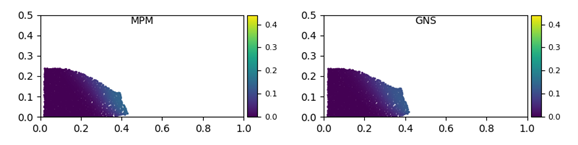}
         \caption{$t/\tau_c=1.5$}
         \label{fig:short3}
     \end{subfigure}
          \begin{subfigure}[b]{0.8\textwidth}
         \centering
         \includegraphics[width=\textwidth]{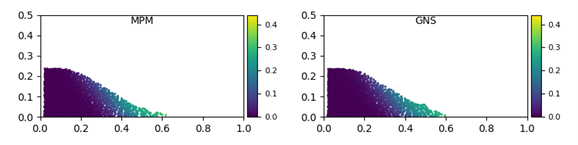}
         \caption{$t/\tau_c=2.5$}
         \label{fig:short4}
     \end{subfigure}
          \begin{subfigure}[b]{0.8\textwidth}
         \centering
         \includegraphics[width=\textwidth]{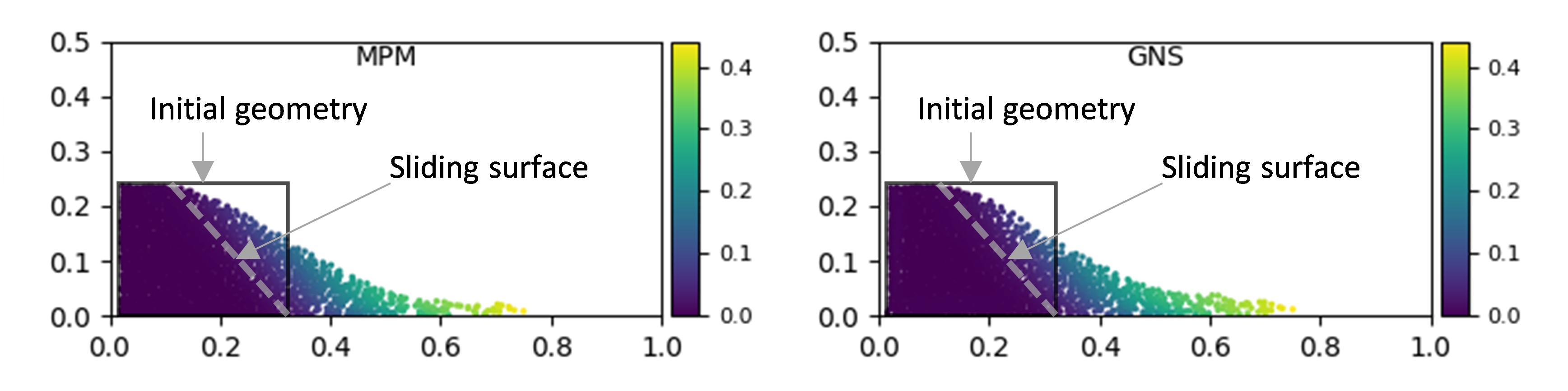}
         \caption{$t/\tau_c=6.4$}
         \label{fig:short5}
     \end{subfigure}
        \caption{Evolution of flow with normalized time for GNS and MPM for the short column with $a=0.8$. Units are in $m$. The color represents the magnitude of the displacement. Subfigure (e) shows the final deposit at the last timestep.}
        \label{fig:short_column}
\end{figure}

We now evaluate the GNS rollout (prediction) of the granular flow dynamics with time for a short column ($a=0.8$). \Cref{fig:short_column} shows the time evolution of granular flow for the short column collapse. We use a normalized time ($t/\tau_c$) to compare the flow evolution, where $t$ is physical time, and  $\tau_c$ is the critical time defined as the time required for the flow to fully mobilize. $\tau_c$ is defined as $\sqrt{H_0/g}$, where $g$ is the gravitational acceleration. In \cref{fig:short_column}, the collapse shows three stages. First, the flow is mobilized by the failure of the flank and reaches full mobilization around $t/\tau_c = 1.0$. The majority of the runout occurs until $t/\tau_c = 2.5$. Beyond $t/\tau_c> 2.5$, the spreading decelerates due to the basal friction and finally stops at around $t/\tau_c= 4.0$ for both MPM and GNS rollout (prediction). As seen in \cref{fig:short_column}, although the GNS has only seen an aspect ratio $a = 1.0$ during training, GNS successfully captures the overall time-evolution of granular flows for a short column ($a=0.8$).

\begin{figure}[!htbp]
     \centering
     \begin{subfigure}[b]{0.49\textwidth}
         \centering
         \includegraphics[width=\textwidth]{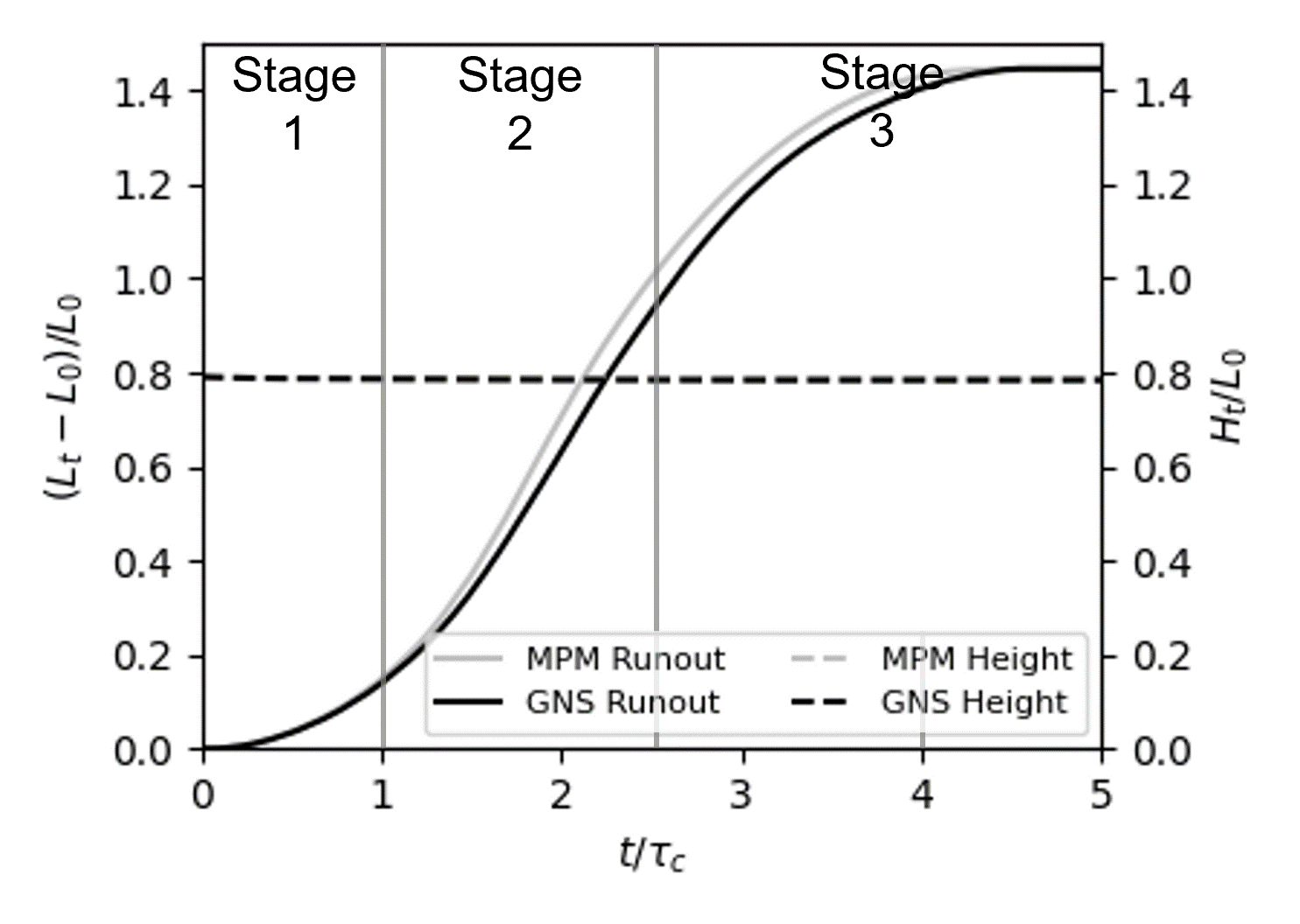}
         \caption{}
     \end{subfigure}
     \hfill
     \begin{subfigure}[b]{0.49\textwidth}
         \centering
         \includegraphics[width=\textwidth]{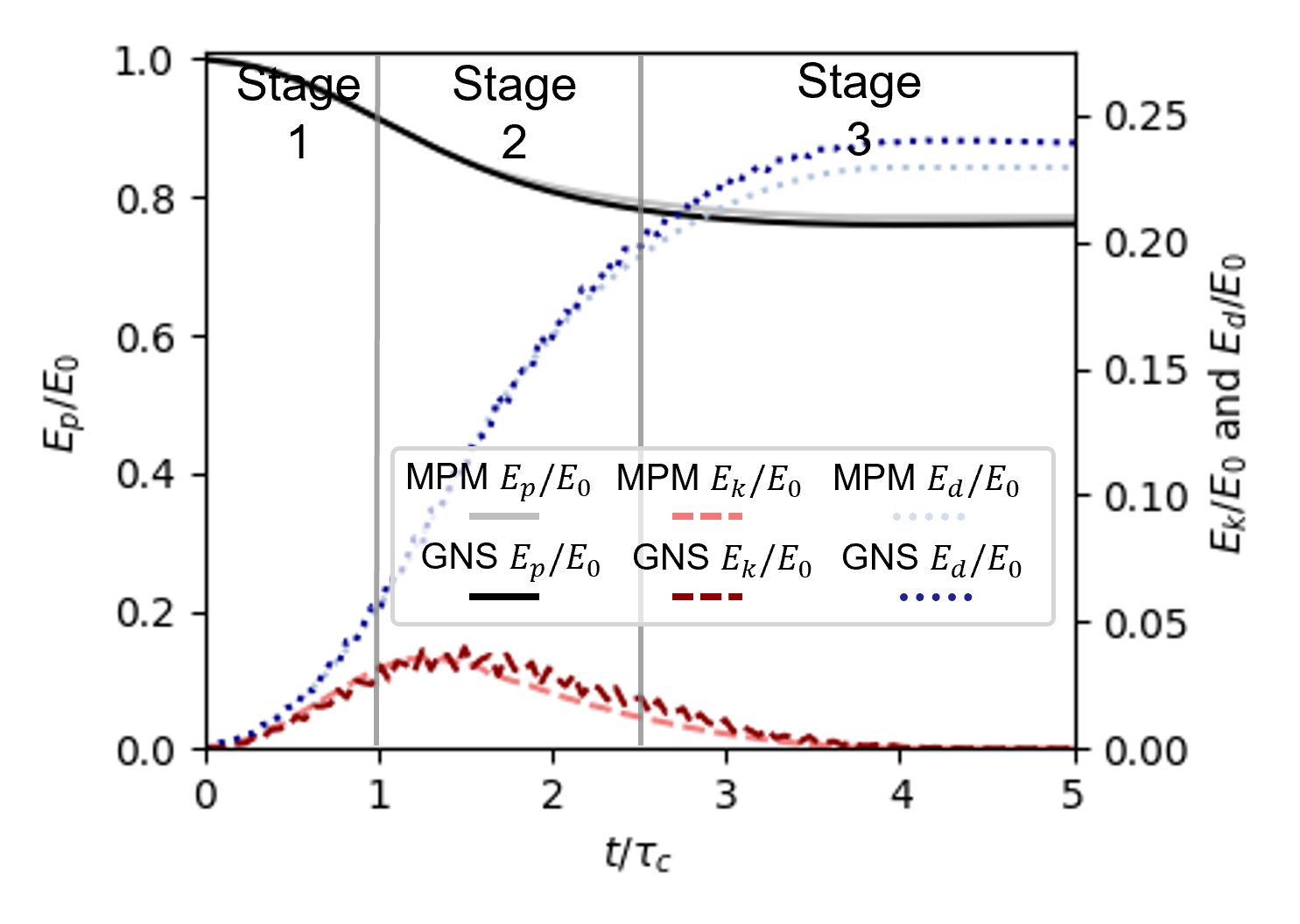}
         \caption{}
     \end{subfigure}
     \hfill
        \caption{(a) Normalized runout and height evolution with normalized time and (b) normalized energy evolution with normalized time for the short column $a=0.8$. $H_t$ is the height from the bottom corner of the boundary to the highest part of the column at $t$. $E_p=\sum_{i=1}^{n}m_igh_i$ is the potential energy of the column, and $E_k=\frac{1}{2}\sum_{i}^{n}m_iv_i^2$ is the kinetic energy of the column, where $m_i$, $h_i$, and $v_i$ is the mass, height, and velocity of the material point $i$, and $n$ is the total number of material points. $E_d=E_0-E_p-E_k$ is the dissipation energy where $E_0$ is the potential energy before material points start to move.}
        \label{fig:short_evolutions}
\end{figure}

In addition to the visual comparison of profiles, we quantitatively investigate the flow dynamics of the GNS rollout of the short column by comparing the normalized runout and height evolution with the MPM. \Cref{fig:short_evolutions}a shows the evolution of normalized runout and height with time. The normalized runout of the MPM (see the gray line corresponding to the left axis in \cref{fig:short_evolutions}a) shows the three stages of collapse. The collapse of the granular column starts with the failure of the flank and evolves slowly until the runout is fully mobilized by $t/\tau_c=1.0$. As the collapse proceeds, the runout acceleration increases ($t/\tau_c$ = 1.0 to 2.5). After this time, the runout deaccelerates due to basal friction, and finally stops at $t/\tau_c\approx4.0$. Both GNS and MPM show a constant normalized height (see the gray line corresponding to the right axis in \cref{fig:short_evolutions}a) as only the flank of the column collapse, leaving a static truncated-conical core. GNS predicts an almost identical evolution of runout as the MPM simulation, which is noteworthy as only a small portion of the training trajectories (5 out of 26) includes the deacceleration behavior leading to the flow coming to rest due to the basal friction before hitting the walls. Overall, the quantitative comparison shown in \cref{fig:short_evolutions}a confirms that the GNS can accurately model the granular flow dynamics for the short column. 

\Cref{fig:short_evolutions}b shows the energy evolutions from GNS rollout and MPM simulation. Based on the principle of energy conservation, the granular flow must satisfy $E_0=E_p+E_k+E_d$, where $E_0$ is the potential energy of the column before material points start to mobilize, $E_p$ is the potential energy, $E_k$ is the kinetic energy, and $E_d$ is the dissipation energy due to friction along the boundary and material. In \cref{fig:short_evolutions}b, both GNS rollout and MPM simulation show identical energy evolutions. A significant fraction of the stored potential energy is converted to kinetic energy in the initial stages of the failure, reaching a peak value of kinetic energy at $t/\tau_c=1$. The kinetic energy dissipates due to the basal friction and flow ceases at $t/\tau_c = 4.0$ when $E_k$ is fully dissipated. 

\subsubsection{Tall column}
\begin{figure}[!htbp]
     \centering
     \begin{subfigure}[b]{0.8\textwidth}
         \centering
         \includegraphics[width=\textwidth]{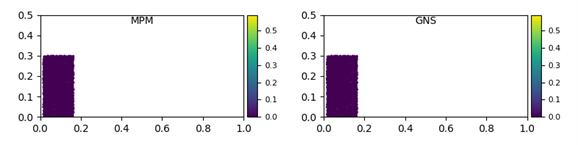}
         \caption{$t/\tau_c=0$}
         \label{fig:tall1}
     \end{subfigure}
     \vfill
     \begin{subfigure}[b]{0.8\textwidth}
         \centering
         \includegraphics[width=\textwidth]{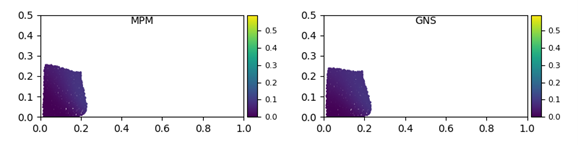}
         \caption{$t/\tau_c=1.0$}
         \label{fig:tall2}
     \end{subfigure}
     \vfill
     \begin{subfigure}[b]{0.8\textwidth}
         \centering
         \includegraphics[width=\textwidth]{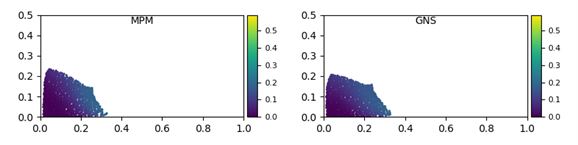}
         \caption{$t/\tau_c=1.5$}
         \label{fig:tall3}
     \end{subfigure}
          \begin{subfigure}[b]{0.8\textwidth}
         \centering
         \includegraphics[width=\textwidth]{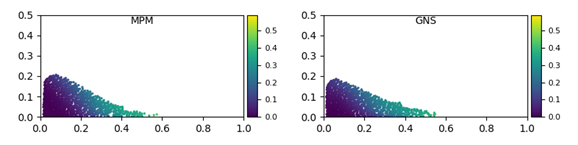}
         \caption{$t/\tau_c=2.5$}
         \label{fig:tall4}
     \end{subfigure}
          \begin{subfigure}[b]{0.8\textwidth}
         \centering
         \includegraphics[width=\textwidth]{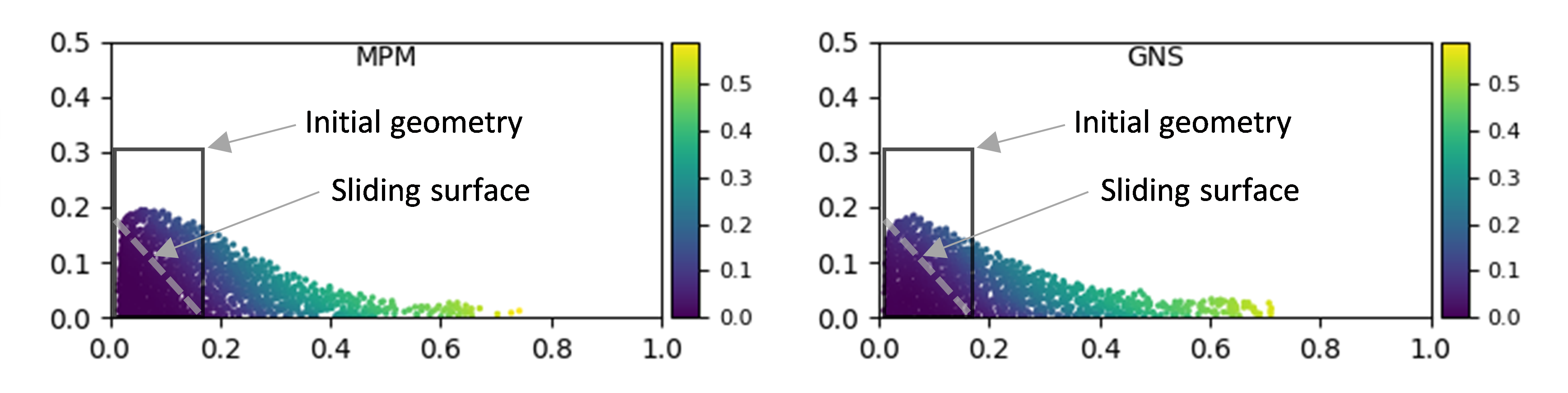}
         \caption{$t/\tau_c=5.7$}
         \label{fig:tall5}
     \end{subfigure}
        \caption{Evolution of flow with normalized time for GNS and MPM for the tall column with $a=2.0$. Units are in $m$. The color represents the magnitude of the displacement. Subfigure (e) shows the final deposit at the last timestep.}
        \label{fig:tall_column}
\end{figure}

Tall columns exhibit different runout dynamics than the short column. GNS was only trained on granular mass with an aspect ratio of 1.0 and has not seen the dynamics of a tall column during training. As an example, we demonstrate the GNS prediction for a tall column with $a=2.0$. \Cref{fig:tall_column} shows the GNS rollout and MPM simulation of the runout evolution for this case. GNS rollout predicts an identical runout profile with normalized time as the MPM simulation. Similar to the short column, the tall column also shows the three stages: the initial mobilization of the flow ($t/\tau_c$ to 1.0), runout ($t/\tau_c$ = 1.0 to 2.5) along the failure surface, deacceleration ($t/\tau_c$ = 2.5 to 4.0). In the tall column, however, a larger volume of sliding mass above the failure plane is mobilized. During the initial stages of the collapse, the granular mass experiences free fall due to gravity dominated by collisional dissipation. As the granular mass reaches the failure surface, the vertical kinetic energy is converted to horizontal acceleration, resulting in longer runout distances. GNS rollout shows similar behavior to the MPM runout simulation.

\begin{figure}[!htbp]
     \centering
     \begin{subfigure}[b]{0.49\textwidth}
         \centering
         \includegraphics[width=\textwidth]{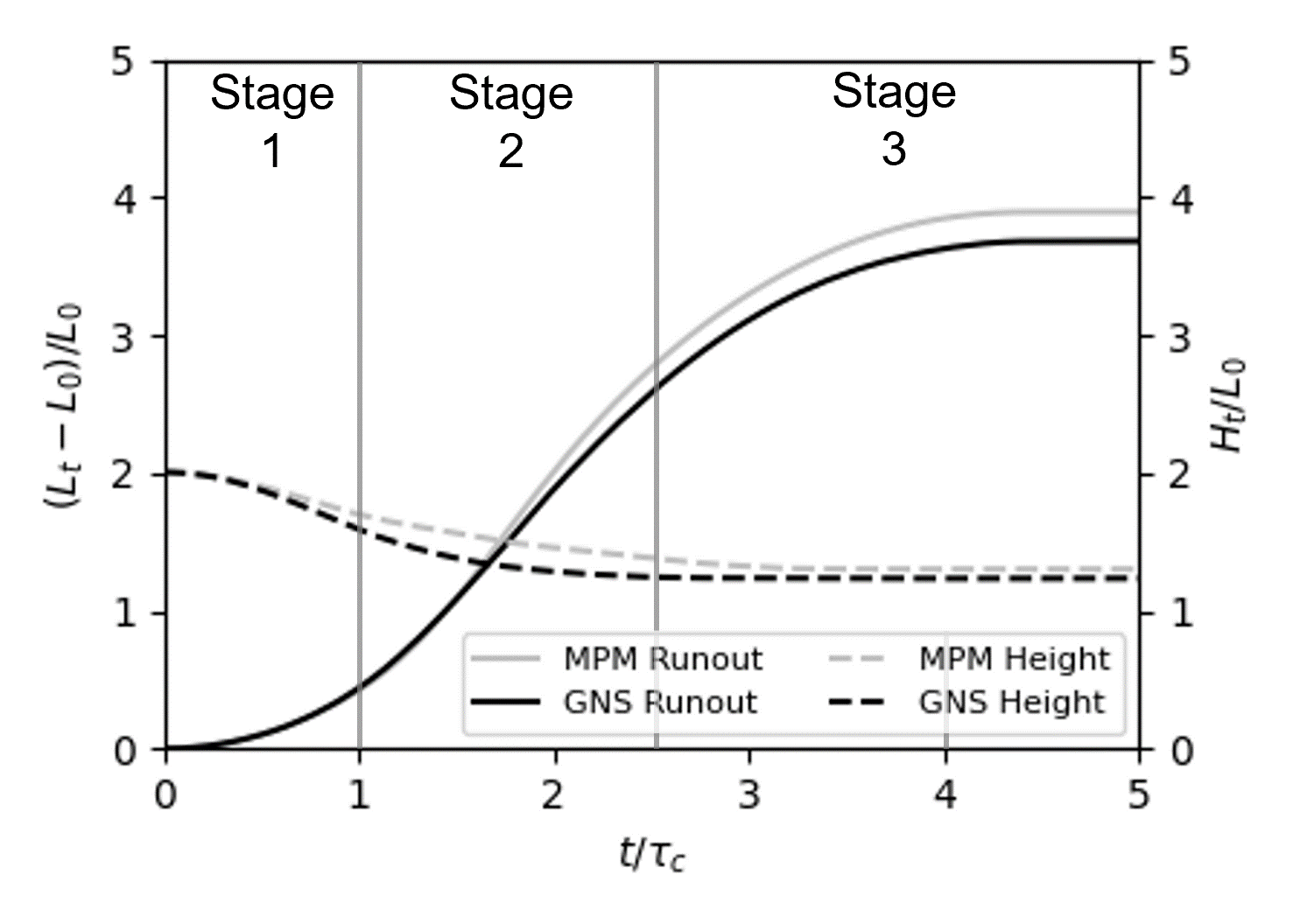}
         \label{fig:tall_runout}
         \caption{}
     \end{subfigure}
     \hfill
     \begin{subfigure}[b]{0.49\textwidth}
         \centering
         \includegraphics[width=\textwidth]{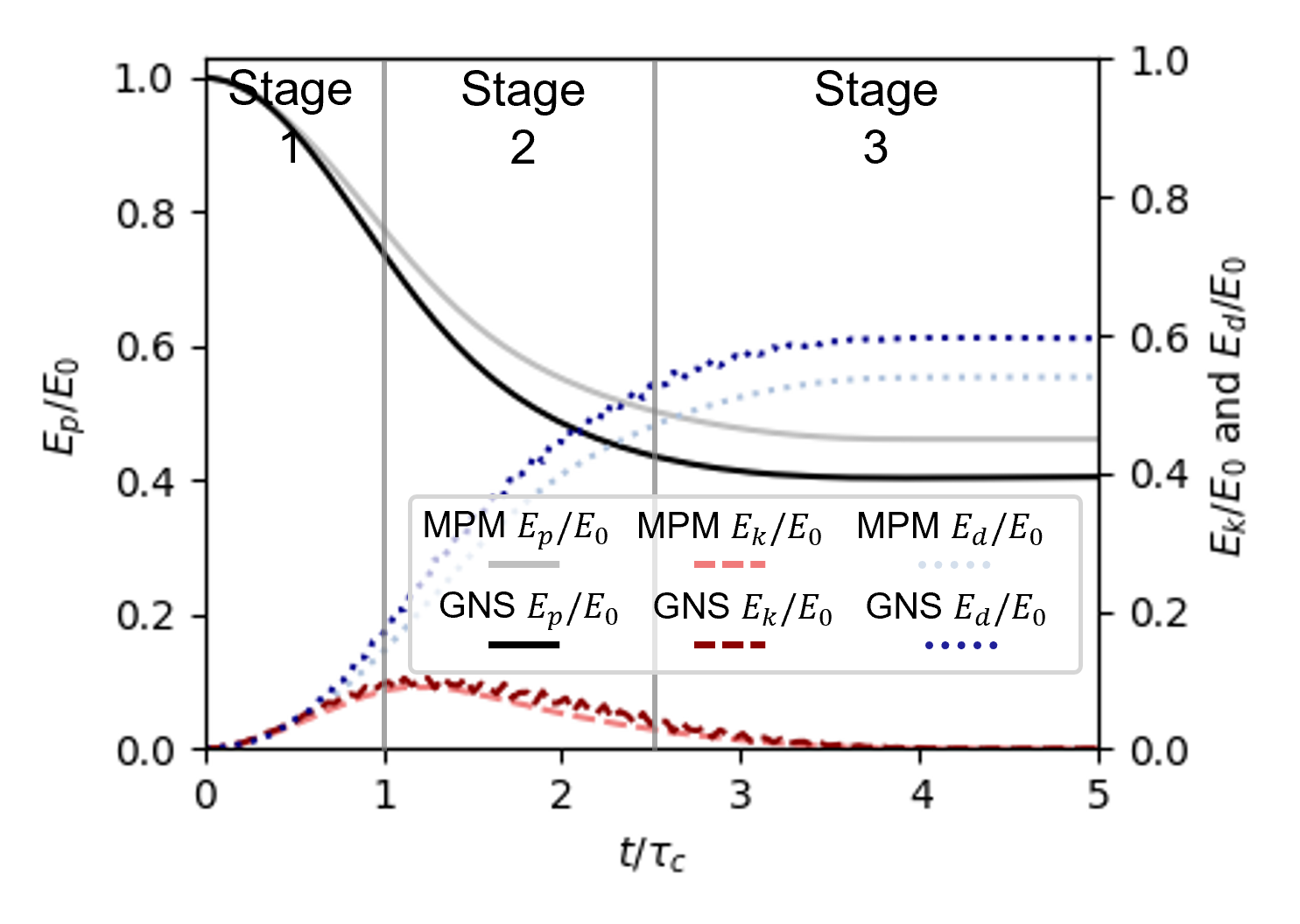}
         \label{fig:tall_energy}
         \caption{}
     \end{subfigure}
     \hfill
        \caption{(a) Normalized runout and height evolution with normalized time and (b) normalized energy evolution with normalized time for the tall column with $a=2.0$.}
        \label{fig:tall_evolutions}
\end{figure}

\Cref{fig:tall_evolutions}a shows the normalized runout and height evolution for the tall column. Although the runout evolution remains identical in the initial phase of the collapse, MPM shows a slightly larger normalized runout compared to the GNS. The final height in both GNS and MPM remains the same. 

\Cref{fig:tall_evolutions}b presents the normalized energy evolution of the GNS rollout and the MPM simulation. During the initial stages of the collapse ($t/\tau_c$ to 1.0), a large amount of initial potential energy is converted to kinetic energy due to the free fall of mass under gravity. Both GNS and MPM show almost identical energy profiles. GNS shows a larger potential energy loss as the flow accelerates with an almost similar gain in kinetic energy. It indicates that GNS predicts larger frictional dissipation in tall columns, which could be from the training data focused only on short columns having higher frictional dissipation than tall columns. At the final stage, MPM shows less dissipation due to the basal boundary friction, resulting in a slightly longer runout than GNS rollout. Generally, energy dissipation behavior in GNS is consistent with MPM showing a more significant potential drop and increased dissipation energy accumulation.

Overall, the GNS rollout is consistent with the MPM simulation with a runout error of 5.7 \% for the tall column with $a=2.0$, implying that the GNS can capture the dynamics of granular flows in collision-dominated tall columns despite only being trained on $a=1.0$. 

\subsubsection{Upscaled domain}
\begin{figure}[!htbp]
     \centering
     \begin{subfigure}[b]{0.8\textwidth}
         \centering
         \includegraphics[width=\textwidth]{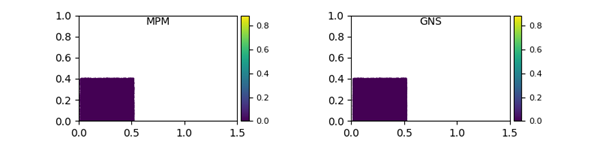}
         \caption{$t/\tau_c=0$}
         \label{fig:upscale1}
     \end{subfigure}
     \vfill
     \begin{subfigure}[b]{0.8\textwidth}
         \centering
         \includegraphics[width=\textwidth]{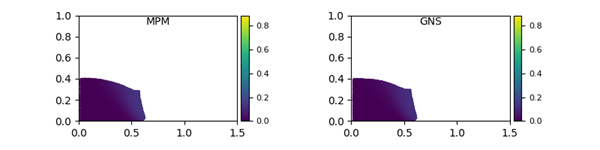}
         \caption{$t/\tau_c=1.0$}
         \label{fig:upscale2}
     \end{subfigure}
     \vfill
     \begin{subfigure}[b]{0.8\textwidth}
         \centering
         \includegraphics[width=\textwidth]{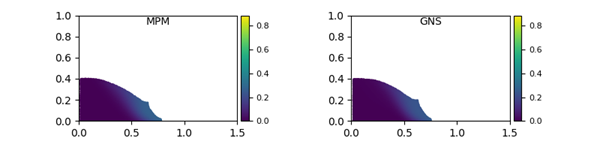}
         \caption{$t/\tau_c=1.5$}
         \label{fig:upscale3}
     \end{subfigure}
          \begin{subfigure}[b]{0.8\textwidth}
         \centering
         \includegraphics[width=\textwidth]{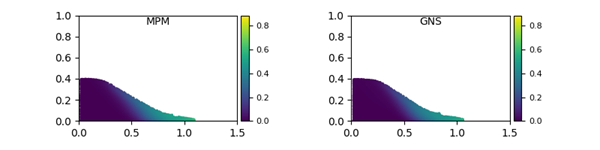}
         \caption{$t/\tau_c=2.5$}
         \label{fig:upscale4}
     \end{subfigure}
          \begin{subfigure}[b]{0.8\textwidth}
         \centering
         \includegraphics[width=\textwidth]{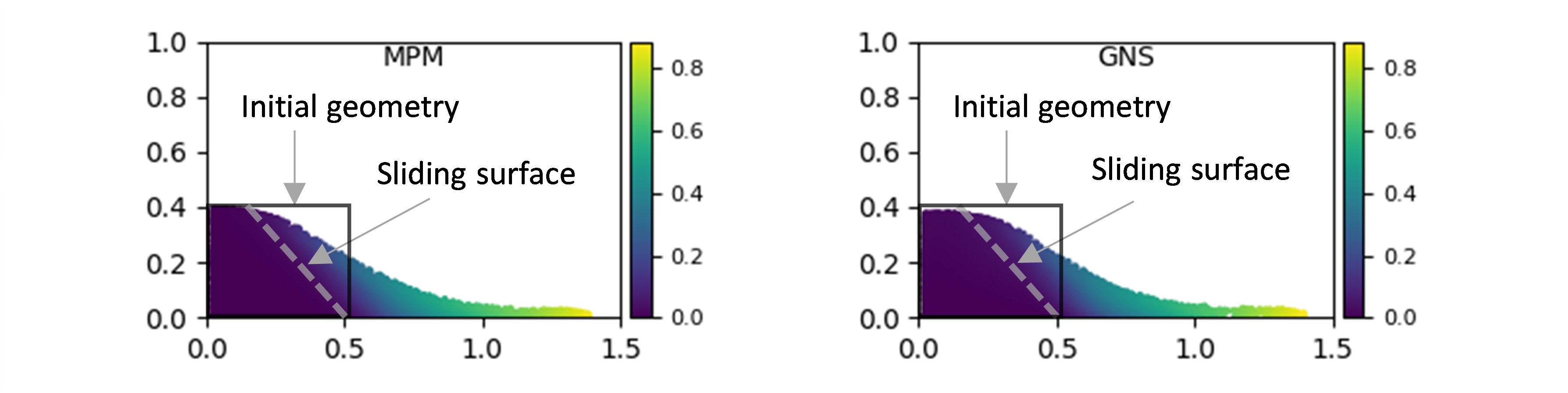}
         \caption{$t/\tau_c=6.4$}
         \label{fig:upscale5}
     \end{subfigure}
        \caption{Evolution of flow with normalized time for GNS and MPM for the upscaled case of short column with $a=0.8$. Units are in $m$. The color represents the magnitude of the displacement. Subfigure (e) shows the final deposit at the last timestep.}
        \label{fig:upscaled_column}
\end{figure}

GNS is generalizable to different initial configurations of the flow simulation owing to the strong inductive bias of the GNN\citep{Battaglia2018}. The strengths of GNS surrogate models would be to train them on small-scale experiments and then predict large-scale dynamic scenarios with complex boundary conditions. We now evaluate the scalability of GNS to a larger domain, including more material points than the training dataset. \Cref{fig:upscaled_column} shows the GNS rollout of a short column $a=0.8$ with 5120 material points (up to 5$\times$ more material points than the training dataset) for a larger simulation domain and longer rollout duration than the training dataset.

\begin{figure}[!htbp]
     \centering
     \begin{subfigure}[b]{0.49\textwidth}
         \centering
         \includegraphics[width=\textwidth]{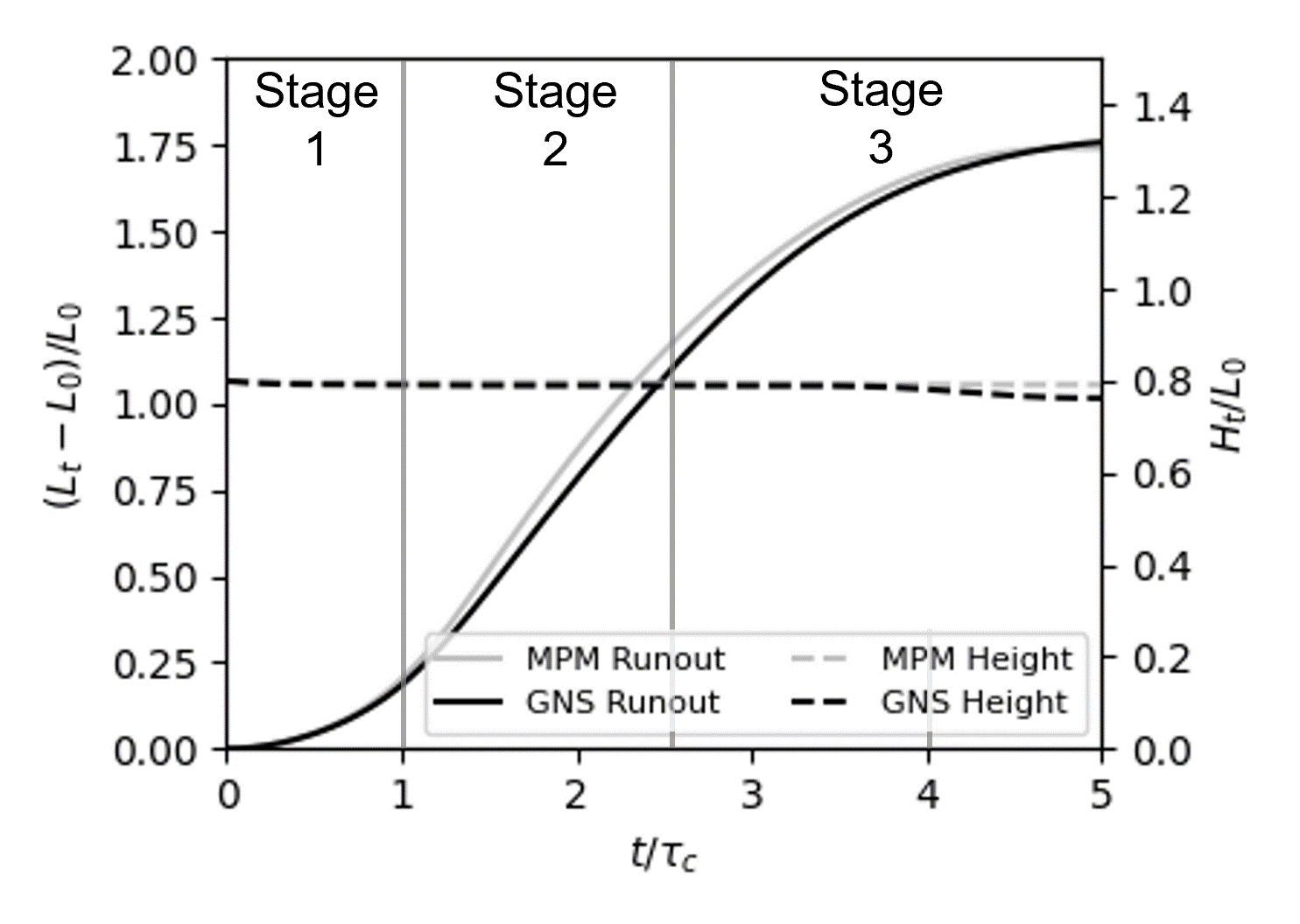}
         \label{fig:upscale_runout}
         \caption{}
     \end{subfigure}
     \hfill
     \begin{subfigure}[b]{0.49\textwidth}
         \centering
         \includegraphics[width=\textwidth]{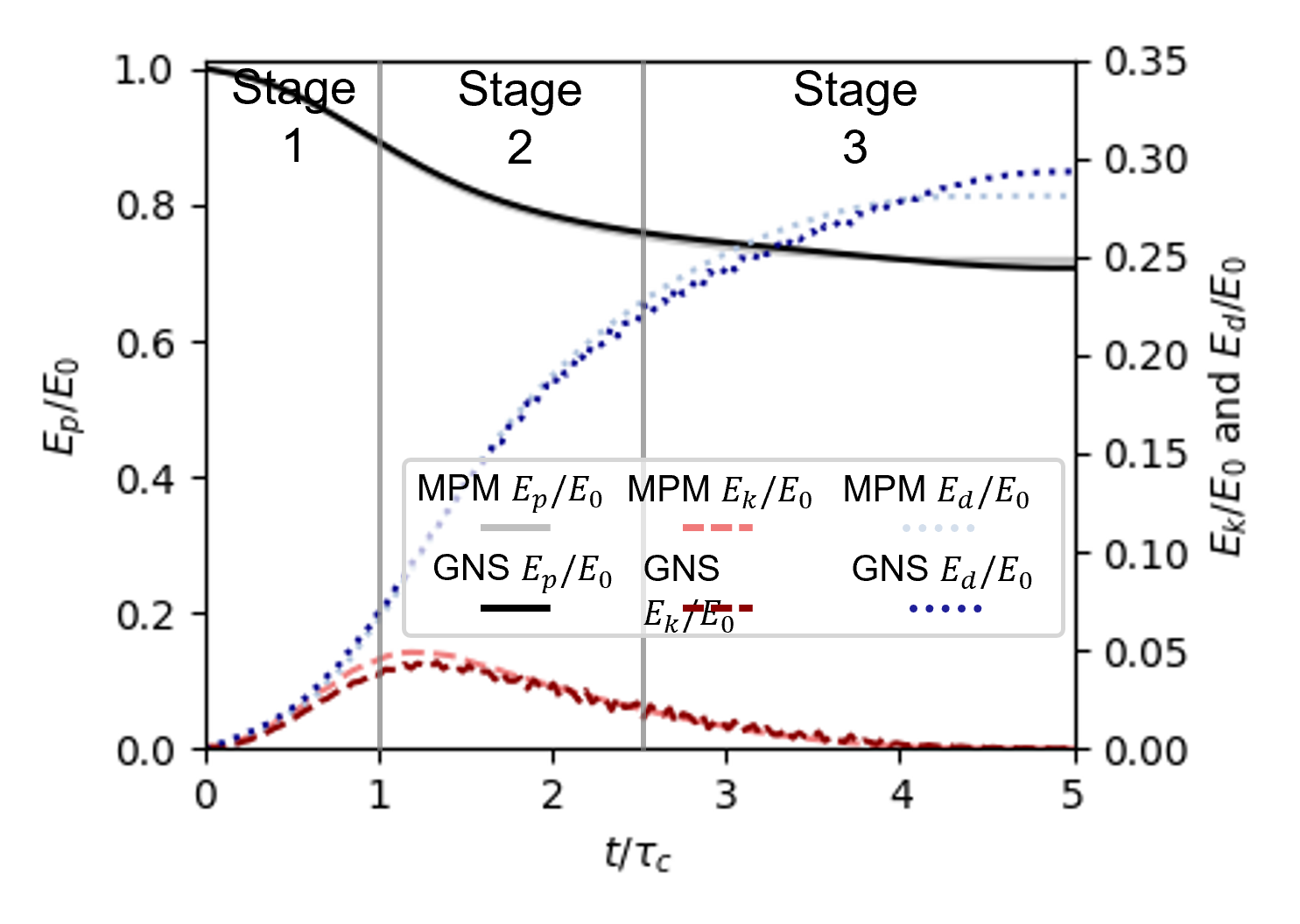}
         \label{fig:upscale_energy}
         \caption{}
     \end{subfigure}
     \hfill
        \caption{Normalized runout and height evolution with normalized time and (b) normalized energy evolution with normalized time for the upscaled case of the short column with $a=0.8$. Note that the data after $t/\tau_c>5.0$ is abbreviated since the flow reaches a static state.}
        \label{fig:upscaled_evolution}
\end{figure}

GNS successfully predicts the flow dynamics for an upscaled domain size showing a similar runout profile with the MPM simulation. The GNS rollout predicts a normalized runout of 1.74 while the MPM simulation shows 1.76, showing an error of 1.30\%. \Cref{fig:upscaled_evolution} shows that GNS rollout successfully replicates energy evolution observed in an upscaled domain compared to the MPM simulation. Hence, GNS can reproduce the flow dynamics even for the upscaled geometries beyond the training dataset.

\begin{figure}[!htbp]
    \centering
    \includegraphics[width=0.7\textwidth]{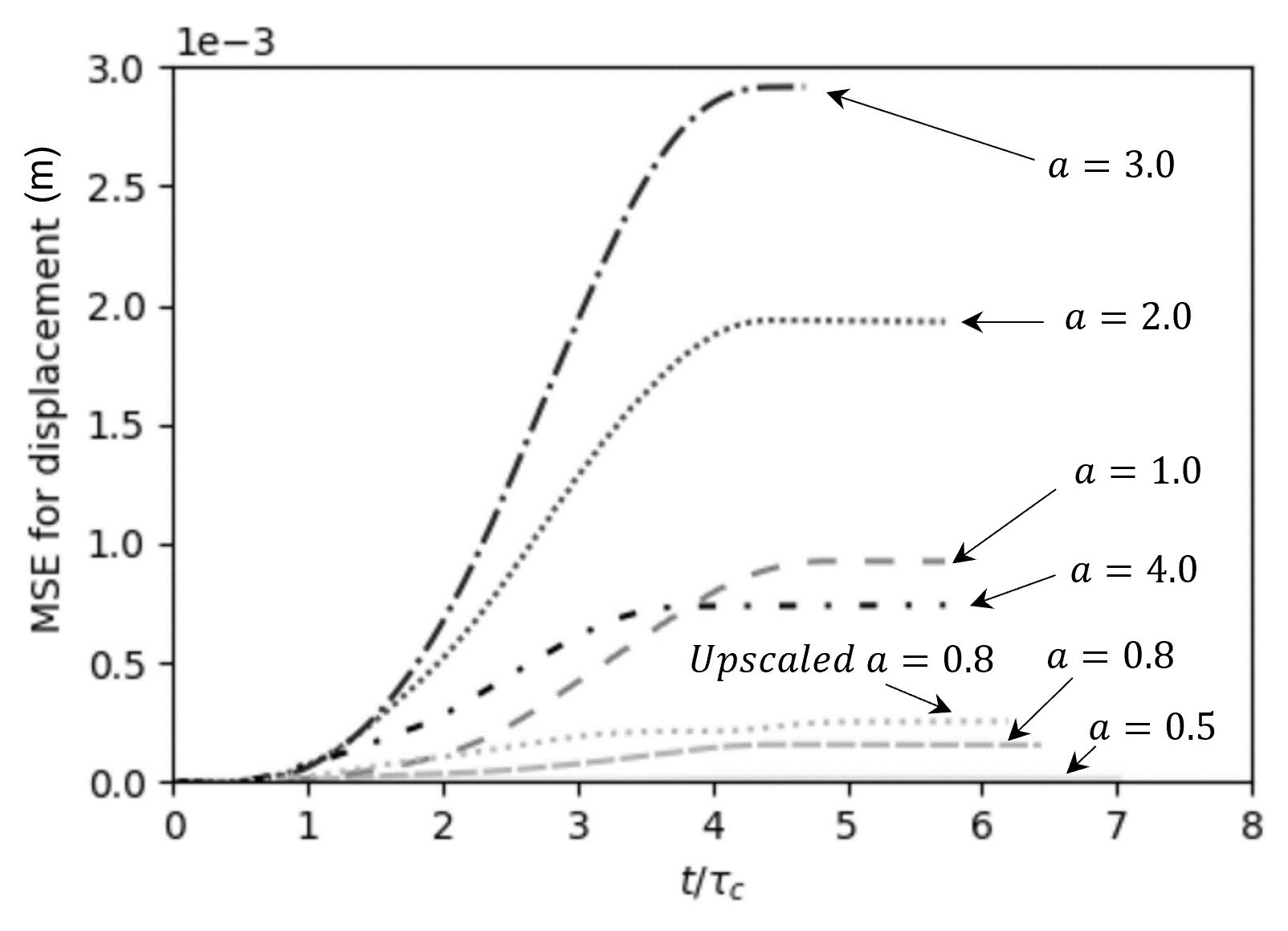}
    \caption{Evolution of mean squared displacement error over all material points with time.}
    \label{fig:error_all_cases}
\end{figure}

\begin{figure}[!htbp]
    \centering
    \includegraphics[width=0.8\textwidth]{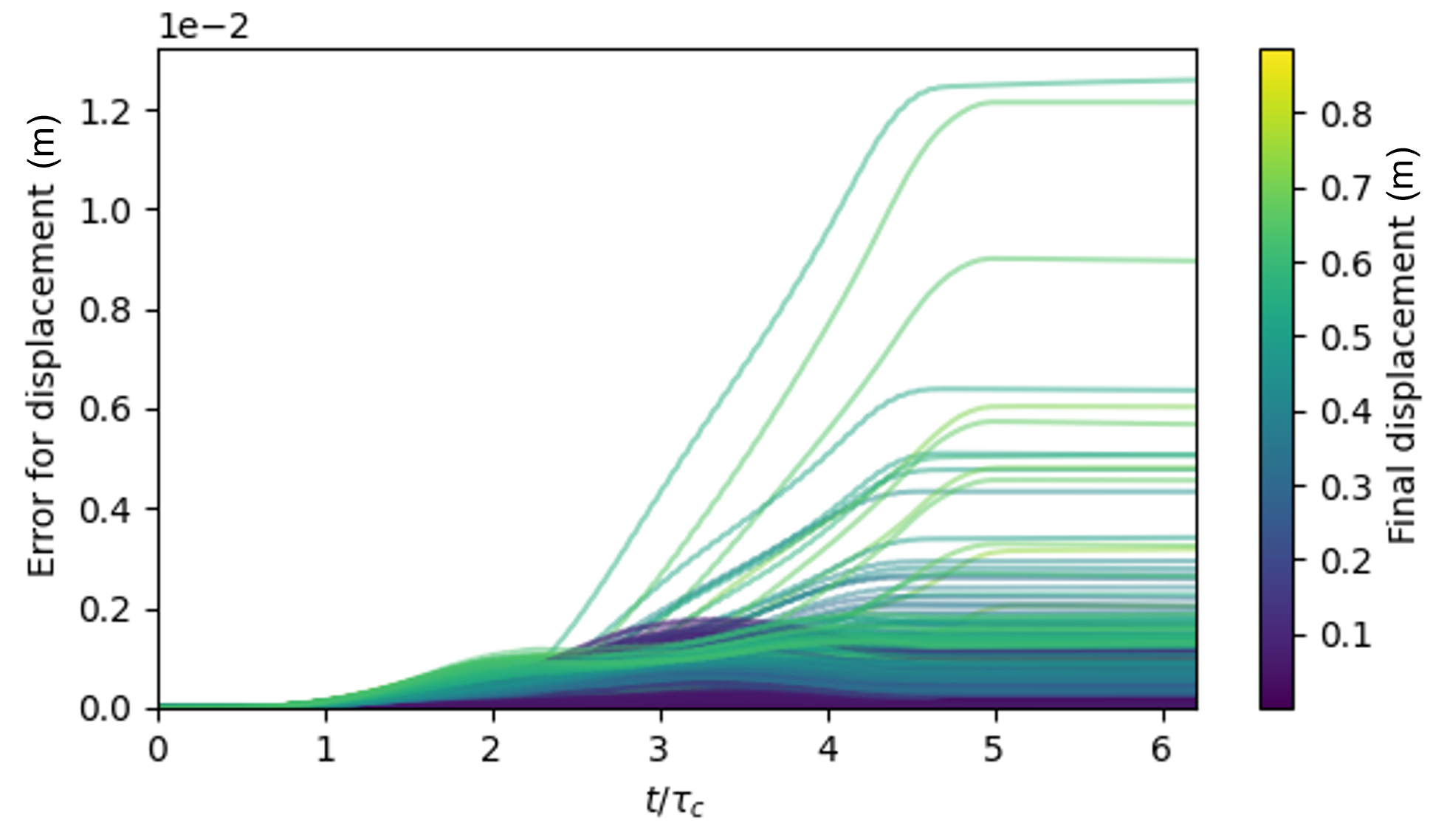}
    \caption{Evolution of squared displacement error for each material point with normalized time in upscaled case of $a=0.8$. The line color represents the final displacement of each material point.}
    \label{fig:error_particles}
\end{figure}

The primary source of GNS rollout error is not from the simulation scale but from the portion of material points that shows a large amount of displacement during column collapse. \Cref{fig:error_all_cases} shows the evolution of mean squared error (MSE) of displacement over all material points ($N$) with time $t$ computed as $\frac{1}{n}\sum_{i}^{N}\left(\boldsymbol{p}_i^t-{\boldsymbol{p}_{MPM}}_i^t\right)^2$, where ${\boldsymbol{p}_{MPM}}_i^t$ is material point position from MPM. When we compare the MSE for $a=0.8$ with 1,824 of material points and its upscaled domain (2.22x material points), upscaling does not alter the MSE significantly. \Cref{fig:error_particles} shows the evolution of the squared error of displacement of individual material points for the upscaled domain ($a=0.8$). The squared error shows larger values for those material points which run out further, i.e., the larger final displacements, but the proportion of the error with respect to the final runout is small so that GNS could simulate the upscaled case without significant error.

\subsubsection{Different friction angles}
\begin{figure}[!htbp]
    \centering
    \includegraphics[width=0.7\textwidth]{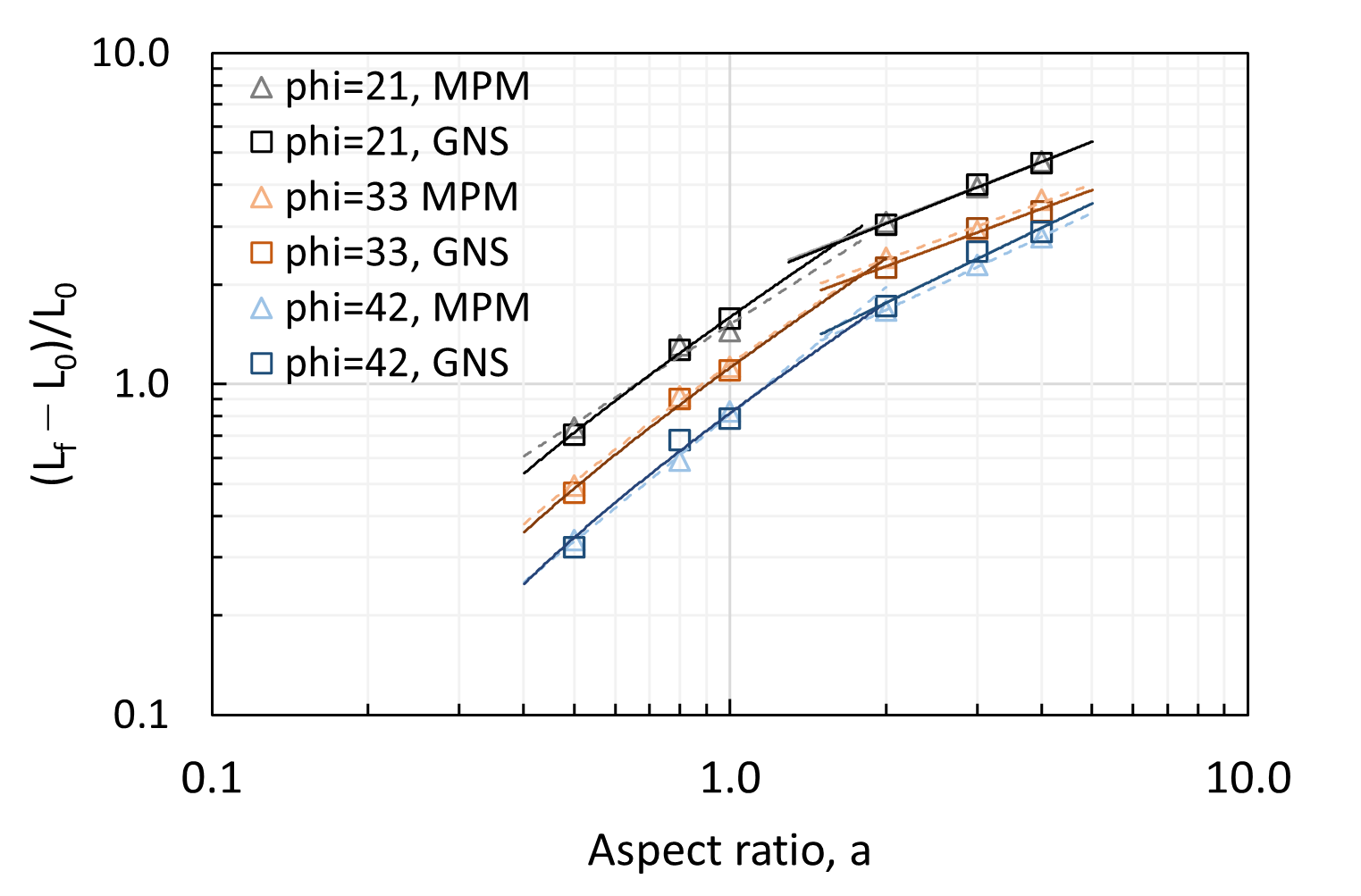}
    \caption{Normalized runout distance ($\left(L_f-L_0\right)/L_0$) with different aspect ratios ($a$) and friction angles ($\phi$).}
    \label{fig:powerlaw_multifriction}
\end{figure}

To make GNS learn the different frictional behaviors, we provided the GNS with the granular flow trajectories with five different friction angles as described in \cref{sec:datasets}. \Cref{fig:powerlaw_multifriction} is the summary of the prediction result. It shows the normalized runout predicted by the GNS compared to MPM on the aspect ratios and friction angles not shown during the training. As we have already observed in the single material case \cref{fig:powerlaw}, the increasing trend of normalized runout with the aspect ratio is well captured through multiple friction angles as well as the relationship transitions between short columns and tall columns. Additionally, GNS generalizes well beyond the geometry (aspect ratio of 1.0) on which it was trained both for short and tall columns. These results suggest that the GNS successfully predicts the different granular flow dynamics depending on various friction angles although an exhaustive range of friction angles is not provided during the training.

\subsection{Granular flow with barriers}\label{sec:result_barrier}
\begin{figure}[!htbp]
    \centering
    \includegraphics[width=1.0\textwidth]{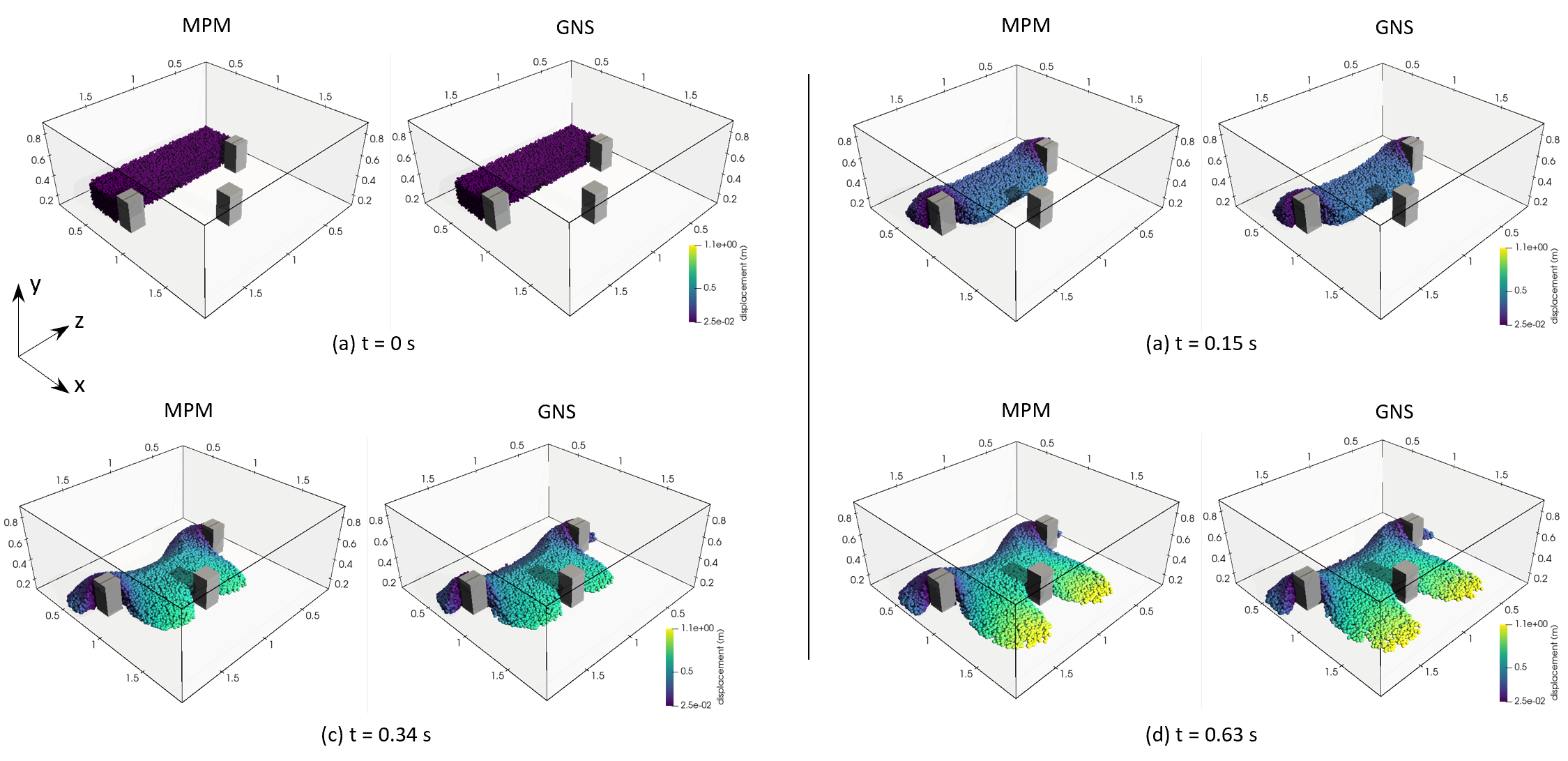}
    \caption{Evolution of flow interacting with baffles for GNS and MPM from initial
condition to the final deposit at the last timestep. The simulation domain is $1.8 \times 0.8 \times 1.8 \ m$ and the initial geometry of the granular mass is $0.35\times0.25\times1.4 \ m$. The barrier size is  $0.15\times0.30\times0.15 \ m$. The center location of the two baffles in the first low are at $(x=0.775, \ z=0.375) \ m$ and $(x=0.775, \ z=1.625) \ m$, and the final baffle is at $(x=1.275, \ z=1.0) \ m$}
    \label{fig:progress-barrier}
\end{figure}

To evaluate the generalizability of GNS for predicting interactions with barriers, we set the simulation configuration of the debris-resisting baffles to be outside the training distribution described in \cref{sec:datasets}. \Cref{fig:progress-barrier}a shows the initial configuration of the simulation. The simulation boundary is from 0.1 to 1.9 for the x-axis, 0.1 to 0.9 for the y-axis, and 0.1 to 1,9 for the x-axis. The granular mass of $0.35\times0.25\times1.4 \ m$ for each dimension approaches three baffles, whose geometry is $0.15\times0.30\times0.15 \ m$, with the initial x-velocity of 2.0 $m/s$. The center location of the two baffles in the first row is at $(x=0.775, \ z=0.375) \ m$ and $(x=0.775, \ z=1.625) \ m$, and that of the final baffle is at $(x=1.275, \ z=1.0) \ m$.  Compared to training data, the simulation domain is four times larger with 2.2 times more material points. While the initial granular mass geometry takes on a cuboid shape, the training data exclusively features cubes. Additionally, we test the GNS with three baffles, in contrast to the training data which only contains one or two.

From the initial state to t=0.15 s (\cref{fig:progress-barrier}), the granular debris propagates downstream uniformly and impacts the first baffle row. Upon hitting the first row, the side of the flow is obstructed by the baffles and the rest of the flow between the baffles proceeds towards the next baffle. Simultaneously, material is deposited and built up upstream of the baffles due to the obstruction and almost reaches the height of the baffle. From t=0.15 s to t=0.34 s, the flow impacts the next baffle and diverges into two granular jets through the open area with roughly symmetric shapes. A smaller amount of material deposition forms upstream of the last baffle compared to the first row. Beyond t=0.34 s, the spreading of the grains decelerates due to the basal friction and finally reaches static equilibrium at around t=0.63 s. GNS rollout successfully replicates the overall kinematics of the flow including the complex interaction with the obstacles. 

\begin{figure}[!htbp]
     \centering
     \begin{subfigure}[b]{0.49\textwidth}
        \centering
        \includegraphics[width=\textwidth]{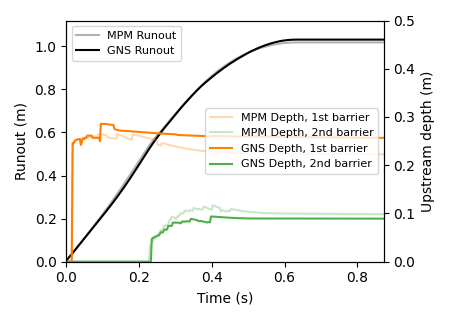}
        \caption{}
        \label{fig:runout-barrier}
     \end{subfigure}
     \hfill
     \begin{subfigure}[b]{0.49\textwidth}
        \centering
        \includegraphics[width=\textwidth]{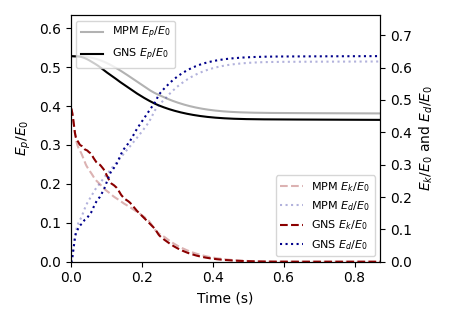}
        \caption{}
        \label{fig:energy-barrier}
     \end{subfigure}
     \hfill
     \caption{(a) Runout and upstream depth evolution and (b) normalized energy evolution with time.}
      \label{fig:barrier_evolution}
\end{figure}

To quantitatively compare the results from GNS and MPM, we measure the runout evolution (\cref{fig:runout-barrier}), similar to the granular column collapse case. In this study, we define runout as the longest travel distance of the flow along the path compared to the initial forefront of the granular mass. With the initiation of the flow, runout rapidly propagates until t=0.34 s experiencing the complex interaction between structures. Subsequently, the runout gradually decelerates and eventually halts as the rest of the flow energy dissipates. GNS perfectly predicts the runout evolution simulated by MPM although the simulation configuration includes more obstacles and a larger domain size compared to the training data.

We also measure the upstream depth evolution of the granular mass deposition right behind the baffles (\cref{fig:runout-barrier}). The upstream flow depth refers to the depth of the granular flow measured upstream of the baffle array when the flow impacts the baffles. As runout proceeds, the flow faces the baffles at the first row and, subsequently, the second baffle. Upon impact with the first baffles, the upstream depth spikes almost to the height of the baffles (0.3 m) due to the deposition of the grains and sudden backwater effects. After peaking, it shows a small amount of decrease over time as the granular flow around the baffles gradually reaches static equilibrium. As the flow impacts the baffle in the next row, the upstream depth surges, and it gradually decreases again but at a lower rate. The overall trend of the upstream depth change is well characterized by GNS, although GNS predicts a slightly less decrease in depth compared to the MPM. However, the difference is not significant.

\Cref{fig:energy-barrier} shows the energy evolution. Initially, the approaching granular flow has high kinetic energy due to the initial potential energy being mobilized. The material accumulation behind the baffles contributes to the steep drop in the kinetic energy, as the flow passes and interacts with the baffles. GNS captures these energy evolution trends accurately. The results from \cref{fig:progress-barrier} and \cref{fig:barrier_evolution} suggest that the GNS successfully learns the granular flow dynamics interacting with obstacles.

\section{Comparative analysis of computational efficiency}

\subsection{Rollout}
\Cref{table:computation_times} shows the computation times for the MPM, GNS with CPU, and GNS with GPU. We investigate four different cases of the number of material points (“small”, “intermediate”, “large” number of material points, and the “larger” case which slightly exceeds GPU memory capacity) for each GNS model (2D and 3D). The MPM simulations were run on 56 cores of Intel Cascade Lake processors with 128 GB memory on TACC Frontera using CB-Geo MPM code, while the GNS rollouts were computed on a single A100 GPU with 40 GB memory on TACC Lonestar 6. 

As the number of material points increases, MPM computation time increases with $O(n \log n)$ at best and $O(n^2)$ at worst \citep{kumar2019scalable}. For GNS, as the number of material points increases, the edge connections linearly increase, which governs the message passing and computation volume of GNS. Accordingly, it leads to the increase of GNS computation time with linear scaling ($O(n)$). When we compare the computation time of GNS with MPM in 2D model, GNS on CPU shows about 3.5 to 6.3× speed up, and GNS on GPU shows about 369 to 740× speed up. For the 3D model, GNS on GPU shows about 15 to 28× speed up, and GNS on GPU shows about 1819 to 2867× speed up. However, a limitation of the GPU-based GNS in its current form is its dependency on the available GPU memory. When the number of material points increases to 102K for 2D and 439K for 3D, the model can't proceed with the rollout due to memory constraints. Despite this, the CPU version of GNS can handle computations for a larger number of material points, surpassing the capability of its GPU counterpart. It's worth noting that the 3D model can simulate a higher number of material points compared to the 2D model because it has fewer edge connections within the connectivity radius.

\cite{arduino2021tsunami} demonstrated that GPU MPM simulations speed up 5 to 10$\times$ compared to the CPU CB-Geo MPM version. Despite this, GNS still provides a great deal of performance boost over GPU versions of numerical simulations, which are also restricted to smaller domains due to GPU memory bandwidths.

GNS can run on multiple GPUs, which increases computational efficiency and the number of simulation scenarios that can be run. Surrogate models, such as GNS, are computationally efficient compared to full-scale numerical simulations with MPM. Therefore, GNS allows for evaluating thousands of scenarios rather than just a handful of cases with full-scale numerical modeling.

\begin{table}
\centering
\caption{Computation time of MPM and GNS rollout per timestep for varying number of material points. Note the 2D model is the one used for simulating granular column collapse, and the 3D model is for the flow with barriers.}
\label{table:computation_times}
\resizebox{\textwidth}{!}{%
\begin{tabular}{cccccccc} 
\toprule
\multirow{2}{*}{Model} & \multirow{2}{*}{\# material points} & MPM & \multicolumn{5}{c}{GNS} \\ 
\cmidrule{3-8}
 &  & \begin{tabular}[c]{@{}c@{}}Time\\(s/timestep)\end{tabular} & \begin{tabular}[c]{@{}c@{}}\# edges\\~(Approx.)\end{tabular} & \begin{tabular}[c]{@{}c@{}}Time-CPU\\(s/timestep)\end{tabular} & Speed up & \begin{tabular}[c]{@{}c@{}}Time-GPU\\(s/timestep)\end{tabular} & Speed up \\ 
\midrule
2D & \begin{tabular}[c]{@{}c@{}}3.6K \\(“small”)\end{tabular} & 37 & 0.153M & 5.9 & 6.3 & 0.050 & 740 \\
 & \begin{tabular}[c]{@{}c@{}}40K \\(“intermediate”)\end{tabular} & 227 & 1.8M & 65.3 & 3.5 & 0.616 & 369 \\
 & \begin{tabular}[c]{@{}c@{}}90K \\(“large”)\end{tabular} & 596 & 4.0M & 148 & 4.0 & 1.515 & 393 \\
 & \begin{tabular}[c]{@{}c@{}}102K \\(“larger”)\end{tabular} & 818 & 4.6M & 170 & 4.8 & Exceeds  memory & \ \\ 
\midrule
3D & \begin{tabular}[c]{@{}c@{}}6.9K \\(“small”)\end{tabular} & 129 & 0.125M & 4.59 & 28.1 & 0.045 & 2867 \\
 & \begin{tabular}[c]{@{}c@{}}250K \\(“intermediate”)\end{tabular} & 2,767 & 4.8M & 174.12 & 15.9 & 1.521 & 1819 \\
 & \begin{tabular}[c]{@{}c@{}}389K \\(“large”)\end{tabular} & 4,149 & 7.2M & 271.56 & 15.3 & 1.996 & 2079 \\
 & \begin{tabular}[c]{@{}c@{}}439K \\(“larger”)\end{tabular} & 4,628 & 8.2M & 306.92 & 15.1 & Exceeds  memory & \ \\
\bottomrule
\end{tabular}%
}
\end{table}

\subsection{Training}
The training cost of GNS is significant, as illustrated by the metrics in \ref{table:training_cost} which include the number of parameters, edge connections, memory usage, and time taken for each training step. The 2D model, which has approximately 20K edge connections, takes 0.26 seconds for each training step. In comparison, the more intricate 3D model, which has about 32K edge connections, takes 0.38 seconds per step. Given the current hyperparameters, roughly 5 million training steps are required to achieve optimal rollout performance. This translates to an estimated training duration of 15 days for the 2D model and 22 days for the 3D model.

We consider using training examples with fewer edge connections to decrease training costs. However, this approach risks compromising our learning capacity. Reducing the connectivity radius decreases edge connections, limiting our model's ability to grasp vital local interactions. Alternatively, when we reduce the number of vertices in the training example, we restrict the range of interactions our model captures. The challenge lies in finding a balance between training efficiency and learning depth. We must ensure the connectivity and number of vertices remain high enough to capture complex interactions yet not so high as to be computationally inefficient. There is a need to balance training costs with optimal learning outcomes.

\begin{table}
\centering
\caption{Training cost of GNS}
\label{table:training_cost}
\resizebox{\textwidth}{!}{
\begin{tabular}{ccccc} 
\toprule
Model & \#
  parameters & \begin{tabular}[c]{@{}c@{}}\# edges \\(approx.)\end{tabular} & \begin{tabular}[c]{@{}c@{}}GPU memory usage \\(approx.) (GB)\end{tabular} & \begin{tabular}[c]{@{}c@{}}Training speed \\(s/step)\end{tabular} \\ \midrule
2D & 1591826 & 20K & 20 & 0.26 \\
3D & 1592979 & 32K & 30 & 0.38 \\
\bottomrule
\end{tabular}}
\end{table}

\section{Limitations of GNS}

The GNS surrogate model demonstrates accuracy, generalizability, and improved computation efficiency. However, the GNS suffers from the following limitations. 

As discussed earlier in \cref{table:computation_times}, the GPU memory limits the current implementation of the GNS surrogate model. A GPU with 40 GB memory can simulate up to around 439K material points (approximately 8.2M edge connections). However, this shortcoming can be improved by optimizing the size of the connectivity radius $R$. For example, in \cref{table:computation_times}, the 3D model is able to simulate 4$\times$ more material points with 2$\times$ more edge connections than the 2D model, since $R$ of the 3D model uses a smaller $R$ (=0.025) and includes less number of material points inside it, while the 2D model uses a larger $R$ (=0.030) and includes more number of material points than the 3D model. Nevertheless, the 3D model still shows an accurate rollout as shown in \cref{sec:result_barrier}. This implies that, although a larger $R$ can consider more interaction between neighbors, excessively large $R$ is not necessary and can aggravate the computation efficiency of GNS. Multi-GPU GNS rollouts will enable the scalability of GNS to larger and more complex domains. 

Another limitation is that GNS shows error accumulation with time, see~\cref{fig:error_all_cases}. GNS uses the explicit Euler time integration to predict the next state based on the current state ($\boldsymbol{X}_t \rightarrow \boldsymbol{X}_{t+1}$) as described in \cref{sec:updater}. GNS uses acceleration to predict the runout of the next time step. However, long duration rollouts result in error accumulation. To overcome this issue, where the GNS predictions could deviate from the reality, we introduce Gaussian noise at each step of the training data to minimize error accumulation. Although including noise effectively reduces error accumulation \citep{Sanchez2020}, GNS can not be free from the error accumulation as timestep increases. The use of purely data-driven architecture, i.e., MLP, for the encoder, processor, and decoder can be a potential source of error as GNS learns to predict acceleration. Substituting MLPs in the encoder and decoder with non-data-driven architectures such as simple linear layers or principal component analysis might help reduce the error of GNS using less training data. \cite{Sanchez2020} investigated the validity of the former method, but confirmed no significant changes in performance. Imposing known physics-based constraints to the message passing in the processor \citep{seo2019differentiable} could also help address potential errors stemming from the purely data-driven architecture. \cite{Yang2022unseen_physics} introduced momentum conservation law to the message passing, and observed the error reduction. 

The calibration of the GNS model is pivotal to ensure the accuracy and reliability of simulations by aligning them with real-world data. Calibration often involves adjusting the model parameters, and in the case of GNS, modifying material properties $\boldsymbol{m}$  or connectivity radius are direct methods. For instance, changing the connectivity radius can affect simulation outcomes, such as granular flow deformation. However, merely tweaking parameters might not suffice for optimal accuracy. An alternative approach is transfer learning, where a GNS model pre-trained on foundation data can be further refined based on experimental observations, enhancing its predictive accuracy for specific scenarios. Continued research into GNS-specific calibration is crucial for its application in real-world engineering analyses.

\section{Conclusion}
Traditional numerical methods are computationally intensive when simulating large-scale granular flows. Statistical or conventional machine learning-based surrogate models are not generalizable since they do not explicitly consider the underlying physics. \added{We implemented a graph neural network (GNN)-based simulator (GNS) as a multi-GPU parallel version for generalizable granular flow surrogate simulation}. The use of graphs efficiently represents the physical state of interacting material points. At the same time, the message passing operation of GNN encourages the neural network to learn the interaction between material points. The expressive power of graphs and message passing that models the interaction between material points allows GNS to accurately predict granular flow dynamics for various conditions, including those not seen during training. We demonstrate the performance of GNS on granular column collapse and granular flow interaction with barriers. GNS precisely simulates different flow dynamics involved in columns for different initial aspect ratios and can also be applied to the upscaled domain with more than 2$\times$  material points with a longer simulation duration than the data provided for training. GNS can predict the granular flow interacting with barriers that include different granular mass and barrier configurations not seen during training. GNS also shows a remarkable speed-up of up to 2900$\times$ computation speed compared to the parallelized CPU version of MPM. The computational efficiency and generalizability of the GNS surrogate can expedite evaluating runout hazards requiring numerous scenarios.

\section*{Data and code availability}
The GNS code~\citep{kumar2022gns} is available under the MIT icense on GitHub (\url{https://github.com/geoelements/gns}). The training dataset and trained models are published under CC-By license on DesignSafe Data Depot~\citep{designsafe-choi-gns}.

\section*{Acknowledgment}
This material is based upon work supported by the National Science Foundation under Grant No.\#2103937. Any opinions, findings, and conclusions or recommendations expressed in this material are those of the author(s) and do not necessarily reflect the views of the National Science Foundation.

The authors acknowledge the Texas Advanced Computing Center (TACC) at The University of Texas at Austin for providing Frontera and Lonestar6 HPC resources to support GNS training (\url{https://www.tacc.utexas.edu}).




\bibliographystyle{elsarticle-harv} 
\bibliography{refs}





\clearpage
\appendix
\section{Training data}
\label{appendix:training_data}
Figure A.1 illustrates the training data used for the Granular Neural Simulation (GNS) model. These images depict 26 trajectories of granular collapses generated through the CB-Geo MPM (Material Point Method) code. Each trajectory shows the dynamic behavior of granular materials used in training the GNS.
\begin{figure}[H]
     \centering
     \begin{subfigure}[b]{0.49\textwidth}
         \centering
         \includegraphics[width=\textwidth]{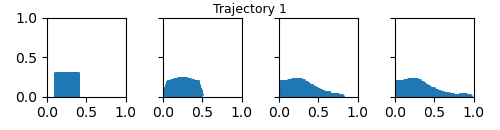}
     \end{subfigure}
     \hfill
     \begin{subfigure}[b]{0.49\textwidth}
         \centering
         \includegraphics[width=\textwidth]{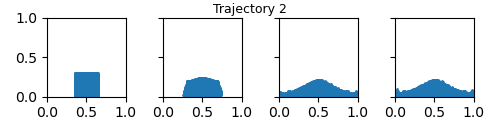}
     \end{subfigure}
\end{figure}

\begin{figure}[H]
     \centering
     \begin{subfigure}{0.49\textwidth}
         \centering
         \includegraphics[width=\textwidth]{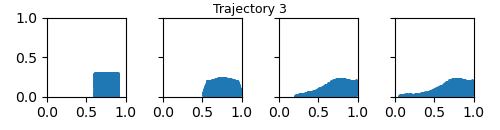}
     \end{subfigure}
     \hfill
     \begin{subfigure}{0.49\textwidth}
         \centering
         \includegraphics[width=\textwidth]{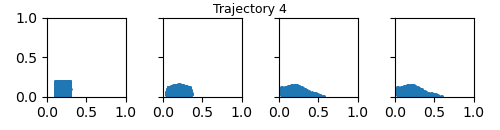}
     \end{subfigure}
\end{figure}

\begin{figure}[H]
     \centering
     \begin{subfigure}{0.49\textwidth}
         \centering
         \includegraphics[width=\textwidth]{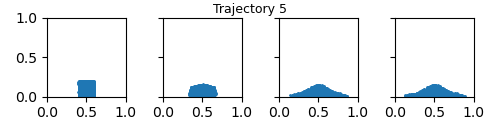}
     \end{subfigure}
     \hfill
     \begin{subfigure}{0.49\textwidth}
         \centering
         \includegraphics[width=\textwidth]{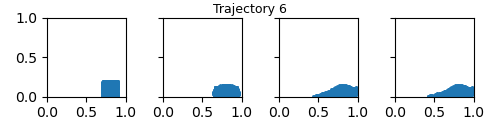}
     \end{subfigure}
\end{figure}

\begin{figure}[H]
     \centering
     \begin{subfigure}{0.49\textwidth}
         \centering
         \includegraphics[width=\textwidth]{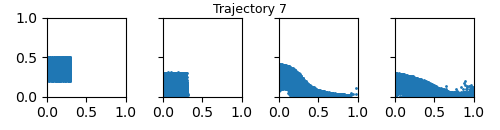}
     \end{subfigure}
     \hfill
     \begin{subfigure}{0.49\textwidth}
         \centering
         \includegraphics[width=\textwidth]{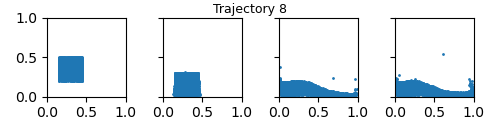}
     \end{subfigure}
\end{figure}

\begin{figure}[H]
     \centering
     \begin{subfigure}{0.49\textwidth}
         \centering
         \includegraphics[width=\textwidth]{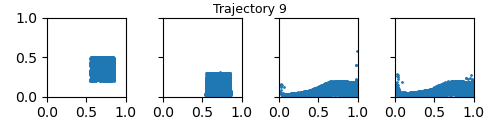}
     \end{subfigure}
     \hfill
     \begin{subfigure}{0.49\textwidth}
         \centering
         \includegraphics[width=\textwidth]{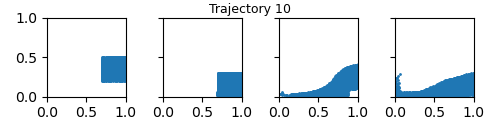}
     \end{subfigure}
\end{figure}

\begin{figure}[H]
     \centering
     \begin{subfigure}{0.49\textwidth}
         \centering
         \includegraphics[width=\textwidth]{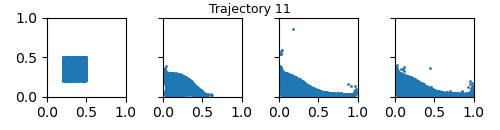}
     \end{subfigure}
     \hfill
     \begin{subfigure}{0.49\textwidth}
         \centering
         \includegraphics[width=\textwidth]{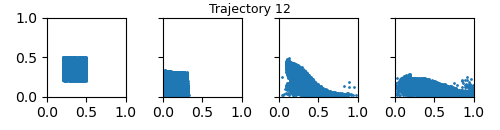}
     \end{subfigure}
\end{figure}

\begin{figure}[H]
     \centering
     \begin{subfigure}{0.49\textwidth}
         \centering
         \includegraphics[width=\textwidth]{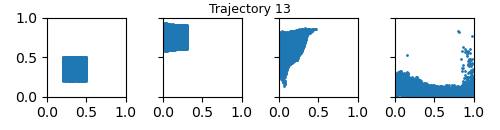}
     \end{subfigure}
     \hfill
     \begin{subfigure}{0.49\textwidth}
         \centering
         \includegraphics[width=\textwidth]{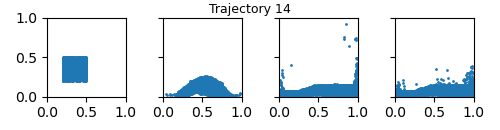}
     \end{subfigure}
\end{figure}

\begin{figure}[H]
     \centering
     \begin{subfigure}{0.49\textwidth}
         \centering
         \includegraphics[width=\textwidth]{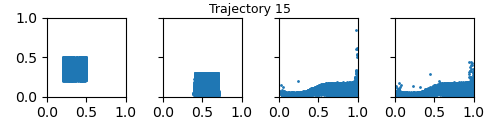}
     \end{subfigure}
     \hfill
     \begin{subfigure}{0.49\textwidth}
         \centering
         \includegraphics[width=\textwidth]{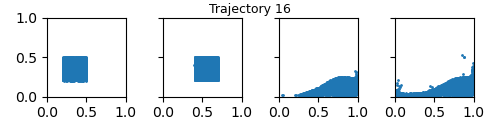}
     \end{subfigure}
\end{figure}

\begin{figure}[H]
     \centering
     \begin{subfigure}{0.49\textwidth}
         \centering
         \includegraphics[width=\textwidth]{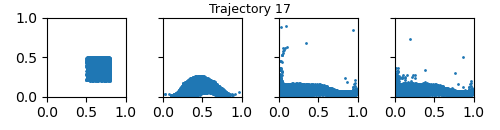}
     \end{subfigure}
     \hfill
     \begin{subfigure}{0.49\textwidth}
         \centering
         \includegraphics[width=\textwidth]{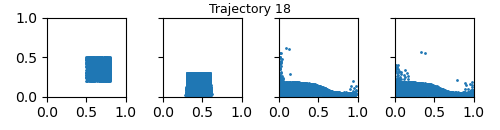}
     \end{subfigure}
\end{figure}

\begin{figure}[H]
     \centering
     \begin{subfigure}{0.49\textwidth}
         \centering
         \includegraphics[width=\textwidth]{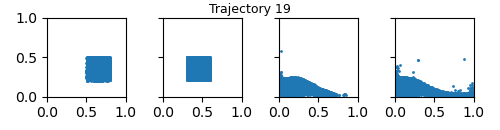}
     \end{subfigure}
     \hfill
     \begin{subfigure}{0.49\textwidth}
         \centering
         \includegraphics[width=\textwidth]{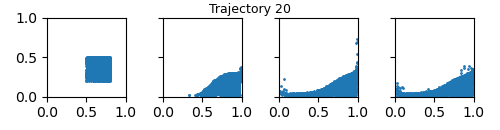}
     \end{subfigure}
\end{figure}

\begin{figure}[H]
     \centering
     \begin{subfigure}{0.49\textwidth}
         \centering
         \includegraphics[width=\textwidth]{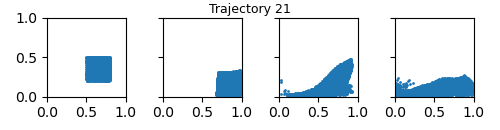}
     \end{subfigure}
     \hfill
     \begin{subfigure}{0.49\textwidth}
         \centering
         \includegraphics[width=\textwidth]{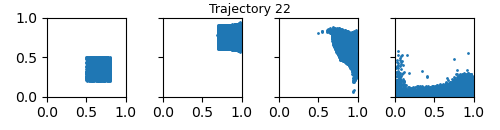}
     \end{subfigure}
\end{figure}

\begin{figure}[H]
     \centering
     \begin{subfigure}{0.49\textwidth}
         \centering
         \includegraphics[width=\textwidth]{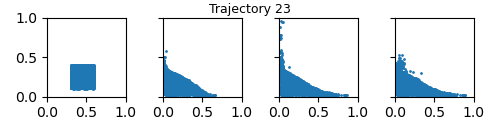}
     \end{subfigure}
     \hfill
     \begin{subfigure}{0.49\textwidth}
         \centering
         \includegraphics[width=\textwidth]{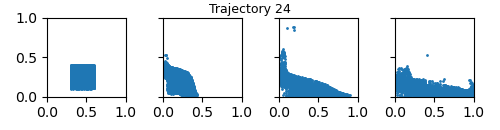}
     \end{subfigure}
\end{figure}

\begin{figure}[H]
     \centering
     \begin{subfigure}{0.49\textwidth}
         \centering
         \includegraphics[width=\textwidth]{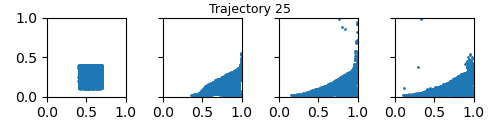}
     \end{subfigure}
     \hfill
     \begin{subfigure}{0.49\textwidth}
         \centering
         \includegraphics[width=\textwidth]{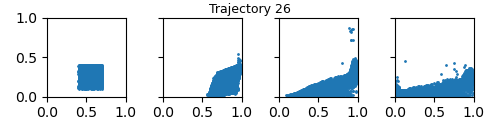}
     \end{subfigure}
\label{fig:training_trajectories}
\caption*{\textbf{Figure A.1.} MPM granular trajectories of training data generated using CB-Geo MPM.}
\end{figure}

\end{document}